\documentclass[12pt]{article}
\pdfoutput=1
\usepackage[english]{babel}
\usepackage[latin1]{inputenc}
\usepackage[normalem]{ulem}
\usepackage{multirow}
\usepackage{booktabs}

\usepackage[text={16.4cm,23.7cm}]{geometry}
\usepackage{soul}
\usepackage{graphicx}
\usepackage[pdftex]{color}
\usepackage{pdfcolmk}
\usepackage{amsmath}
\usepackage{amssymb}
\usepackage{eurosym}
\usepackage{slashed}
\usepackage{hyperref}
\usepackage{wasysym}
\usepackage{feynmp}
\usepackage{float}
\usepackage{placeins}
\usepackage{rotating}
\usepackage{tabularx}

\newcommand{\sub}[2]{#1_{\mathrm{#2}}}
\newcommand{\GeV}{~\mathrm{GeV}}
\newcommand{\TeV}{~\mathrm{TeV}}

\newenvironment{doublecases}
  {\left\lbrace\begin{aligned}}
  {\end{aligned}\right.}
  
\newcommand{\Figref}[1]{Fig.~\ref{#1}}

\newcommand{\Equref}[1]{Eq.~(\ref{#1})}
\newcommand{\EquTworef}[2]{Eqs.~(\ref{#1}) and (\ref{#2})}

\begin{document}

\title{\vspace{-2cm} 
{\normalsize
\flushright TUM-HEP 933/14\\}
\vspace{0.6cm} 
\bf Higher order dark matter annihilations \\ in the Sun and 
implications for IceCube \\[8mm]}

\author{Alejandro Ibarra, Maximilian Totzauer and Sebastian Wild\\[2mm]
{\normalsize\it Physik-Department T30d, Technische Universit\"at M\"unchen,}\\[-0.05cm]
{\it\normalsize James-Franck-Stra\ss{}e, 85748 Garching, Germany}
}

\maketitle
      
\begin{abstract}
Dark matter particles captured in the Sun would annihilate producing a neutrino flux that could be detected at the Earth. In some channels, however, the neutrino flux lies in the MeV range and is thus undetectable at IceCube, namely when the dark matter particles annihilate into $e^+e^-$, $\mu^+\mu^-$ or light quarks. On the other hand, the same interaction that mediates the annihilations into light fermions also leads, via higher order effects, to the production of weak gauge bosons (and in the case of quarks also gluons) that generate a high energy neutrino flux potentially observable at IceCube. 
We consider in this paper tree level annihilations into a fermion-antifermion pair with the associated emission of one gauge boson and one loop annihilations into two gauge bosons, and we calculate the limits on the scattering cross section of dark matter particles with protons in scenarios where the dark matter particle couples to electrons, muons or light quarks from the non-observation of an excess of neutrino events in the direction of the Sun. We find that the limits on the spin-dependent scattering cross section are, for some scenarios, stronger than the limits from direct detection experiments.
\end{abstract}

\section{Introduction}
\label{sec:introduction}

The detection of a high energy neutrino signal from the Sun would strongly point towards exotic physics, most notably the annihilation of dark matter (DM) particles that have been captured in the solar interior. Conversely, the non-observation of a significant excess of events in neutrino telescopes in the direction of the Sun already sets fairly strong limits on the dark matter properties. For example, under the assumption that dark matter captures and annihilations are in equilibrium in the interior of the Sun, present measurements of the neutrino flux by the IceCube experiment set the 90\% C.L. upper limit on the spin-dependent DM-proton interaction cross section $\sigma^{\rm SD}<1.3\times 10^{-40}~(6.0\times 10^{-39})~{\rm cm}^2$ for a 250 GeV dark matter particle that annihilates into  $WW$ ($b\bar b$) \cite{Aartsen:2012kia}. 
Remarkably, for annihilation channels producing hard neutrinos (such as $WW$ or $\tau\tau$), the IceCube limits on the spin-dependent interaction cross section are more stringent than those reported by the most sensitive current experiments probing the same interaction, COUPP \cite{Behnke:2012ys} and SIMPLE \cite{Felizardo:2011uw}.

For some annihilation channels, however, the limits are much weaker, for instance when the dark matter annihilates into $e^+ e^-$, $\mu^+\mu^-$ or $q \bar q$, with $q=u,d,s$. Due to the high density of matter in the interior of the Sun, where the annihilations take place, muons and light hadrons (such as pions and kaons) are quickly stopped before they decay. Therefore, the annihilation into muons or into light quarks produces neutrinos with an energy smaller than $\sim 100$ MeV. Besides, annihilations into electrons do not produce neutrinos directly, although the interaction of the energetic electrons with the solar matter produces pions which  are in turn stopped before decaying, again producing low energetic neutrinos. All these annihilation channels then produce a neutrino flux with an energy well below the detection threshold of IceCube, although they could be detected at SuperKamiokande~\cite{Bernal:2012qh,Rott:2012qb}.

On the other hand, in scenarios where the dark matter particle annihilates into a fermion-antifermion pair, final states with gauge bosons necessarily occur either from internal bremsstrahlung or from loop induced annihilations. Weak gauge bosons can decay producing energetic neutrinos while gluons produce heavy hadrons, which decay also producing energetic neutrinos, hence opening the possibility of observing in IceCube dark matter signals from the Sun in scenarios where the annihilation is driven by couplings to the electron, the muon or the light quarks. In this paper we will investigate this possibility, and derive limits on this class of scenarios from considering the high energy neutrino flux generated by the higher order processes of internal bremsstrahlung or loop annihilations. 

While the annihilation cross sections for the higher order processes are expected to be small, under the common assumption that dark matter captures and annihilations are in equilibrium in the solar interior, the neutrino flux from the Sun is determined, not by the annihilation cross sections, but by the capture rate and by the branching fractions of the different annihilation channels. We will argue that, for some scenarios, the branching fraction for the higher order annihilation processes with gauge bosons in the final state is sufficiently large to produce a high energy neutrino flux at the reach of IceCube, for values of the spin-dependent interaction cross section in agreement with the limits set by the direct search experiments COUPP and SIMPLE. 

The paper is organized as follows. In Section \ref{sec:Contact_Interactions} we undertake a model independent analysis where the tree level dark matter annihilation into a fermion-antifermion is mediated by a contact interaction, and consider separately the limits on the dark matter interaction cross section with protons in scenarios where the most important source of high energy neutrinos is either the annihilation into a fermion-antifermion pair with the associated emission of a gauge boson or the one loop annihilation into two gauge bosons. In Section \ref{sec:toy-model} we focus on a toy model with a Majorana dark matter particle that couples to a light fermion via a Yukawa coupling with a scalar, and where the branching ratios for the two-to-three and the one loop annihilations are calculable in terms of the parameters of the model. Lastly, in Section \ref{sec:Conclusions} we present our conclusions. 
We also include appendices including fitting formulas for the (anti-)neutrino spectra generated by the final state radiation of a gauge boson, tables with the upper limits on the scattering cross section assuming annihilations into $ZZ$ and $gg$ for various dark matter masses, and the expressions for the annihilation cross sections in the relevant channels in the toy model discussed in Section \ref{sec:toy-model}.

\section{Contact interactions}
\label{sec:Contact_Interactions}

\begin{figure}[t]
  \centering
  \includegraphics[width=0.2\textwidth]{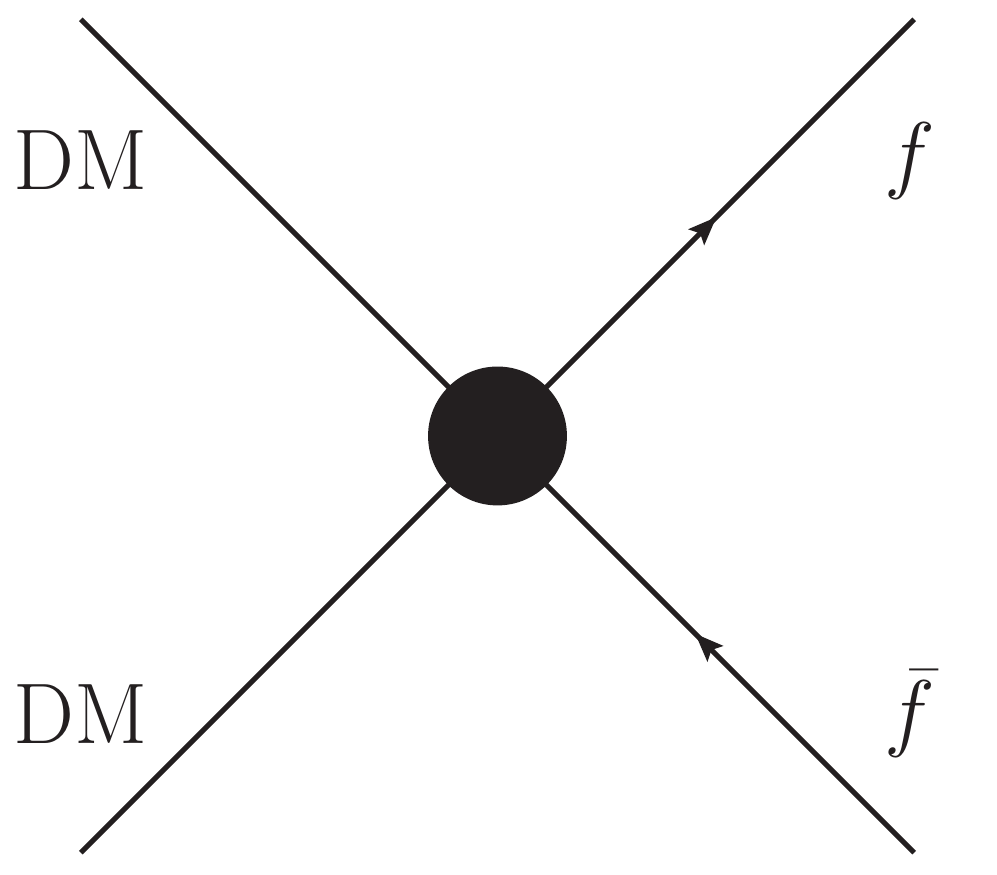}\hspace{1cm}
  \includegraphics[width=0.2\textwidth]{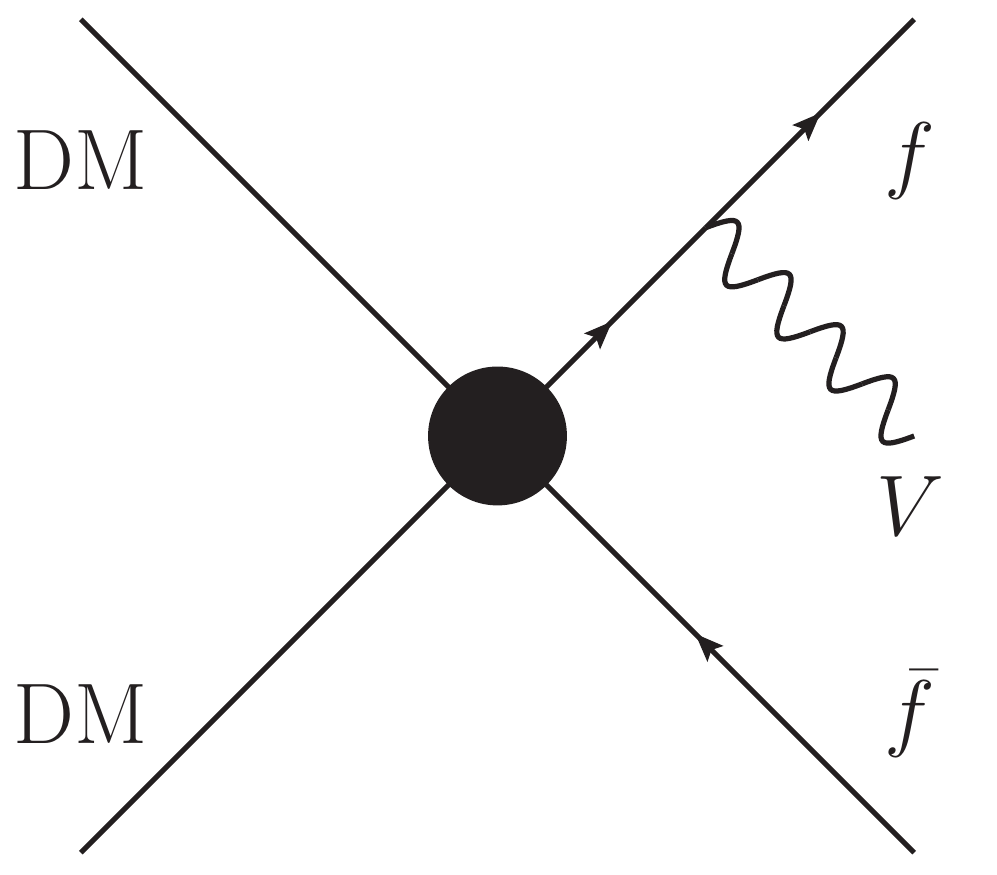}\hspace{1cm}
  \includegraphics[width=0.2\textwidth]{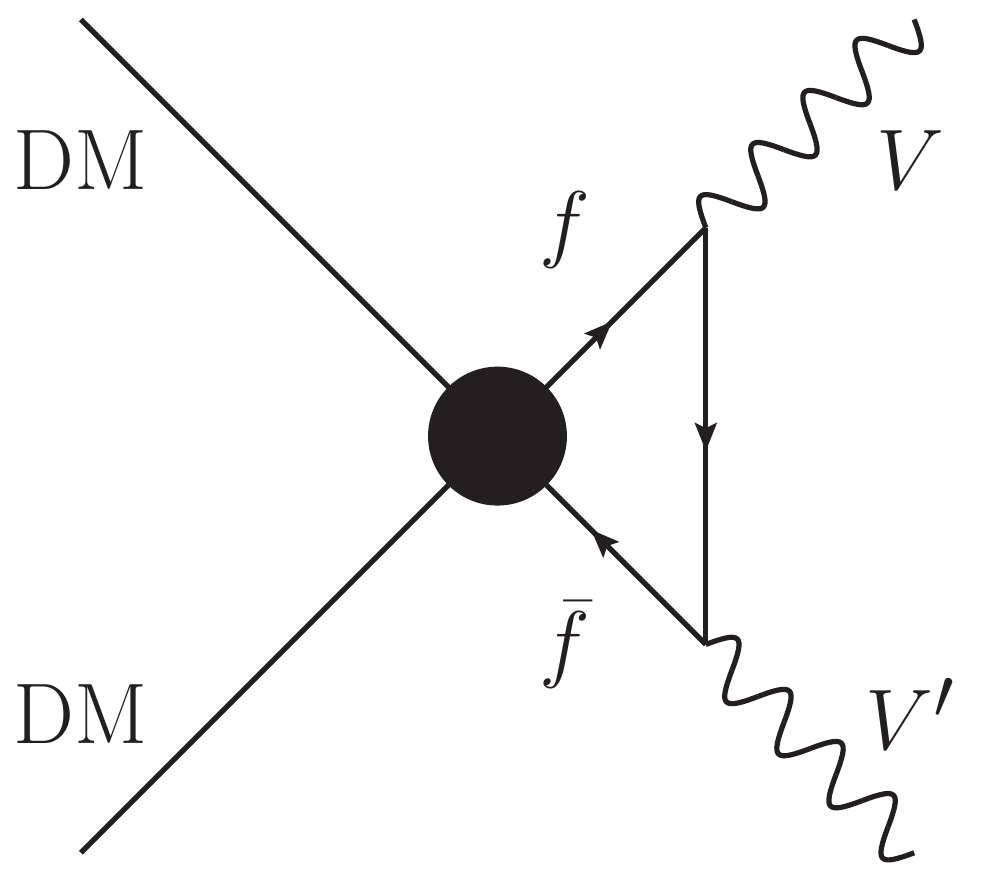}
  \caption{\small Feynman diagrams for the two-to-two annihilation into a fermion-antifermion pair via a contact interaction (left), as well as for the two-to-three annihilation with the associated emission of a gauge boson $V$ (center) and the one-loop annihilation into two gauge bosons  $V$ and $V'$ (right). The last two diagrams contribute to the high energy neutrino flux from the Sun from the decay or the hadronization of the gauge boson in the final state.}
  \label{fig:AnnihilationContactInteraction}
\end{figure}

We consider in this section scenarios where the dark matter particle annihilates into a fermion-antifermion pair, ${\rm DM}\,{\rm DM}\rightarrow f\bar f$, with $f$ an electron, a muon or a light quark, under the assumption that the particles that mediate the interaction are very heavy, such that the annihilation can be described by a contact interaction (see \Figref{fig:AnnihilationContactInteraction}, left plot). We  also assume that the dark matter particle is a $SU(2)_L$ singlet, hence the tree level annihilations into $W^+W^-$ or $ZZ$, which necessarily occur for larger $SU(2)_L$ representations and which would dominate the high energy neutrino flux, cannot take place. 

As discussed in the introduction, dark matter annihilations into electrons, muons or light quarks occurring in the interior of the Sun will not generate a high energy neutrino flux at the Earth. On the other hand, all the Standard Model fermions interact with the $Z$ boson. Therefore, if the dark matter particle annihilates into $f \bar f$, the higher order annihilation process into $f \bar f Z$ will necessarily occur, provided it is kinematically accessible, from the radiation of a soft $Z$ boson off the final fermion~\cite{Kachelriess:2009zy,Ciafaloni:2010ti} (see \Figref{fig:AnnihilationContactInteraction}, middle plot). If the fermion is a $SU(2)_L$ doublet, the annihilation into $f \bar f' W^{\pm}$ is also possible, with $f$ and $f'$ the fermions forming the $SU(2)_L$ doublet. Note that under the assumption that the dark matter particle is a $SU(2)_L$ singlet (and therefore carries no hypercharge as required by the electric charge neutrality) the weak gauge boson cannot be emitted from the initial state but just from the final state. 
Finally, if the fermion in the final state is a quark, the higher order process into  $f \bar f g$ will also occur. Furthermore, the same contact interaction that induces the annihilations into $f\bar f$  necessarily induces the annihilation into $\gamma\gamma$, $\gamma Z$ and $Z Z$ at the one loop level; for left-handed fermions the annihilation into $W^+ W^-$ also occurs, and for colored fermions, the annihilation into $g g$ (see \Figref{fig:AnnihilationContactInteraction}, right plot). In this section we will investigate the possibility of probing in IceCube scenarios where the dark matter annihilation is driven by a coupling to the electron, the muon or a light quark, from the high energy neutrino flux produced in the decay or the hadronization of the gauge bosons produced by these two higher order annihilation processes.

The differential neutrino flux from the annihilation of dark matter particles captured in the interior of the Sun reads:
\begin{equation}
\frac{d\Phi_\nu}{dE_\nu}=\frac{\Gamma_{A}}{4\pi d^2}\sum_i {\rm BF}_i\frac{dN^i_\nu}{dE_\nu}\;,
\label{eq:nuflux}
\end{equation}
where $d=1.5\times 10^{11}\,{\rm m}$ is the distance between the Sun and the Earth, $\Gamma_A$ is the total annihilation rate and $dN^i_\nu/dE_\nu$ is the energy spectrum of neutrinos produced in the annihilation channel $i$ with branching ratio ${\rm BF}_i$. Under the common assumption that dark matter captures and annihilations are in equilibrium in the interior of the Sun, the annihilation rate reads~\cite{Griest:1986yu}:
  \begin{equation}
     \Gamma_A =\frac{1}{2} \Gamma_C \;,
\label{eq:SolutionDMDensitySun}
  \end{equation}
where $\Gamma_C$ is the capture rate, which can be calculated from the scattering cross section of the process dark matter-nucleon and relies on assumptions on the density and velocity distributions of dark matter particles in the Solar System, as well as on the composition and density distribution of the interior of the Sun~\cite{Gould:1987ir}.  In our numerical analysis we have used DarkSUSY \cite{Gondolo:2004sc} to determine the capture rate, adopting the value $v_0 = 270 \, \mathrm{km}/\mathrm{s}$ for the 3D velocity dispersion of the homogeneous Maxwell-Boltzmann dark matter distribution \cite{Green:2011bv} and $\sub{\rho}{local} = 0.3 \GeV/{\rm cm}^3$ for the local dark matter density. For $m_{\mathrm{DM}} \gtrsim 1 $ TeV, we find that the capture rate is approximately given by:
\begin{equation}
\Gamma_C=10^{20} \,{\rm s}^{-1} \left( \frac{1 \, \mathrm{ TeV}}{m_{\mathrm{DM}}} \right)^2 
\frac{2.77 \, \sigma_{\mathrm{SD}} + 4.27 \cdot 10^3 \sigma_{\mathrm{SI}}}{10^{-40} \, \mathrm{cm^2}}\;.
\label{eq:CaptureRate} 
\end{equation}

When equilibration between captures and annihilations is attained in the Sun, the high energy neutrino flux depends critically on the branching ratios of the annihilation channels producing hard neutrinos (see \Equref{eq:nuflux}). The relevant annihilation processes can be classified in three groups: {\it i)} the two-to-two annihilation into a fermion-antifermion pair, {\it ii)} the two-to-three annihilation with the associated production of a gauge boson from final state radiation and {\it iii)} the two-to-two annihilation into gauge bosons via loops. It is important to note that in the contact interaction limit the weak gauge boson in the process {\it ii)} can only be emitted from the final state, and hence the rate for the two-to-three annihilation is proportional to the rate for the two-to-two annihilation into a fermion-antifermion pair. 
Then, there are only two independent groups of processes, namely {\it i)} and {\it iii)}, hence allowing to identify two possible scenarios, depending on whether the rate for the two-to-two annihilation into fermions is much larger or much smaller than the rate for the loop annihilations. 

\subsection{Loop annihilations subdominant}
\label{sec:loop-subdominant}

This scenario arises when the dark matter particle annihilates dominantly into a fermion-antifermion pair, as is the case, for example, when the dark matter particle is a Dirac fermion that annihilates into $f_L {\bar f_L}$ or  $f_R {\bar f_R}$, or a Majorana fermion that annihilates into $f_L {\bar f_R}$ or $f_R {\bar f_L}$ (in both cases, the annihilation proceeds in the s-wave with neither helicity nor velocity suppressions). Hence, the two-to-three annihilation with the associated emission of a gauge boson will have a larger rate than the loop annihilation and will be the most important source of high-energy neutrinos, since the former process is suppressed with respect to the annihilations into a fermion-antifermion pair only by two powers of the weak coupling constant, while the latter by four.

The radiation of gauge bosons from the final state can be described by a set of parton distribution functions $D_{f \rightarrow V} \left( x \right)$ which only depend on the quantum numbers and spin of the final state particle. $D_{f \rightarrow V} \left( x \right)$ can be interpreted as the probability of radiating a gauge boson, $V=Z,W,g$, off a final state fermion, $f$, with a fraction $x$ of the energy of the fermion, $E_V = x E_f$. For a fermion with electric charge $q_f$ and weak isospin $T_{3,f}$ the relevant parton distribution functions are \cite{Ciafaloni:2010ti,Ciafaloni:2001mu,Ciafaloni:2005fm}
\begin{align}
D_{f \rightarrow Z} \left( x \right) = \frac{\alpha_{\text{em}}  \, (T_{3,f} - q_f \sin^2 \theta_{\text{W}})^2}{2\pi \sin^2 \theta_{\text{W}} \cos^2 \theta_{\text{W}}} P_{f \rightarrow V} \left( x \right)\;,
\label{eqn:DfZ}
\end{align}
\begin{align}
D_{f \rightarrow W} \left( x \right) = \frac{\alpha_{\text{em}}\, T_{3,f}^2 }{\pi \sin^2 \theta_{\text{W}}} P_{f \rightarrow V} \left( x \right)\;,
\label{eqn:DfW}
\end{align}
and, if the fermion is a quark, also~\cite{Altarelli:1977zs} 
\begin{align}
D_{f \rightarrow g} \left( x \right) = \frac{\alpha_\text{s}}{2 \pi}  P_{f \rightarrow V}
\label{eqn:Dfg}
\end{align}
where $P_{f \rightarrow V}$ are splitting functions given by
\begin{align}
P_{f \rightarrow V} \left( x \right) = 
\begin{cases}
&\displaystyle{\frac{1+\left(1-x\right)^2}{x} \left[ \ln \frac{x^2 m_{\rm DM}^2}{M_V^2} + 2 \ln \left( 1+\sqrt{1-\frac{M_V^2}{x^2 m_{\rm DM}^2}} \right)\right]}\hspace{0.3cm}{\rm for~} V=Z,W \,, \vspace{0.4cm}\\
&  \displaystyle{\frac{4}{3}\frac{1+(1-x)^2}{x}} \hspace{0.3cm} {\rm for~} V=g\,.
\end{cases}
\label{eqn:PF}
\end{align} 
For our analysis we use PYTHIA 8.176 \cite{Sjostrand:2006za,Sjostrand:2007gs} which includes the emission of weak gauge bosons and gluons from the final state, as described in detail in \cite{Christiansen:2014kba}.\footnote{Our resulting neutrino spectra at Earth are, for $m_{\rm DM} \gtrsim 300$ GeV, in good agreement with those presented in \cite{Baratella:2013fya}.}

The spectrum of neutrinos originating from the final state radiation is then modified in order to take into account the effect of the dense medium where the annihilation takes place, similarly as in \cite{Ibarra:2013eba}. Namely, muons and light hadrons are stopped, while the energy loss and subsequent decay in flight of taus and heavy hadrons were simulated following the chain of scatterings they undergo inside the Sun. Lastly, the neutrino propagation from the solar interior to the surface and eventually to the Earth was calculated using the Monte Carlo code WimpSim \cite{Blennow:2007tw} adopting the most recent best fit values for the neutrino oscillation parameters~\cite{GonzalezGarcia:2012sz}. We include in Appendix \ref{ap:parametrization} fitting functions for the (anti-)neutrino spectra at the Earth produced in various annihilation channels including the final state radiation of weak gauge bosons and (for quarks) gluons.

Finally, we calculate the induced number of (anti-)muon events in IceCube following the approach of \cite{Scott:2012mq}, using the effective area presented in \cite{DanningerPhD}, and derive an upper limit on the spin-independent and spin-dependent interaction cross sections with protons (assuming isospin invariance) from the non-observation in IceCube-79 of an excess of events with respect to the expectations from the atmospheric background. To optimize the limits we choose for each dark matter mass the cut angle between the reconstructed muon direction and the Sun that gives the best constraint under a background only hypothesis, as described in the Appendix of \cite{Ibarra:2013eba}. 

\begin{figure}[h!]
\begin{center}
\includegraphics[width=0.49\textwidth]{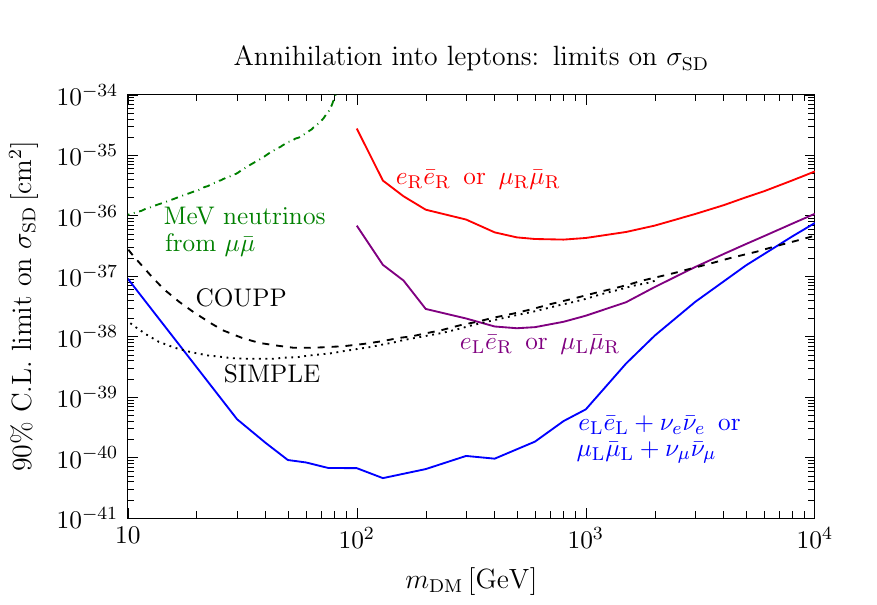}
\includegraphics[width=0.49\textwidth]{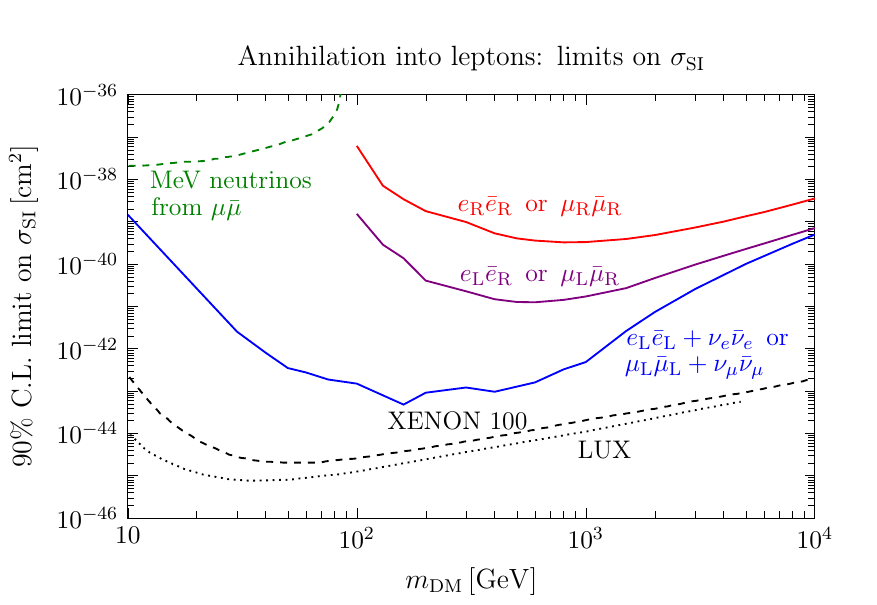} \\
\includegraphics[width=0.49\textwidth]{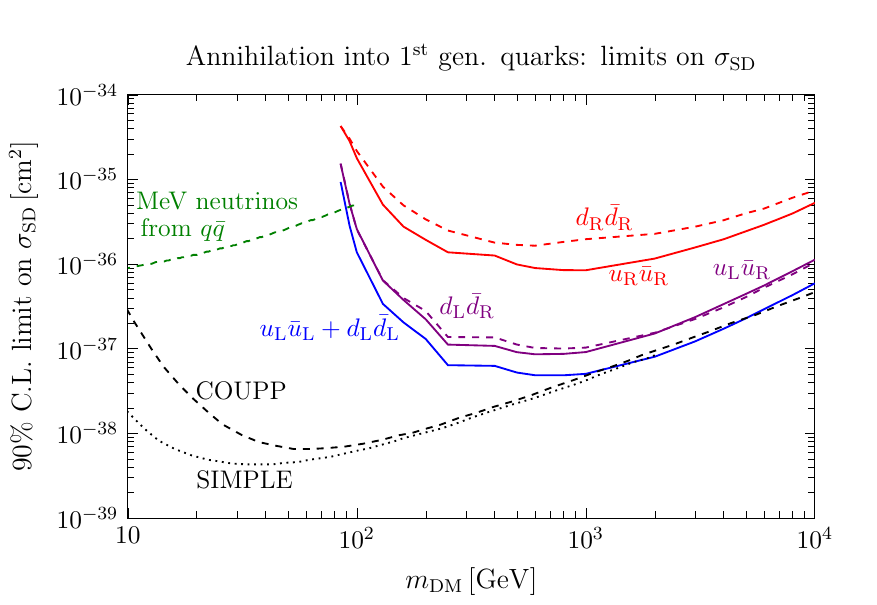}
\includegraphics[width=0.49\textwidth]{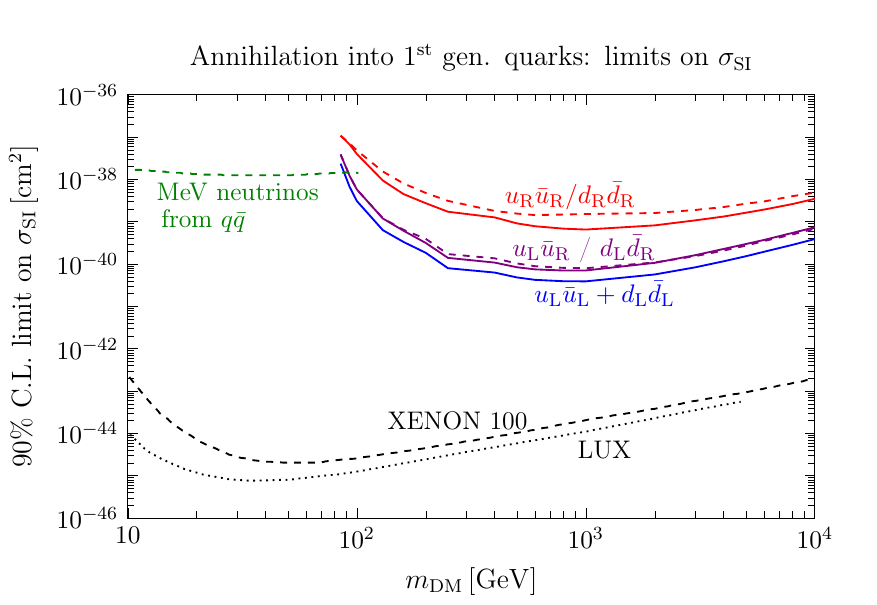} \\
\includegraphics[width=0.49\textwidth]{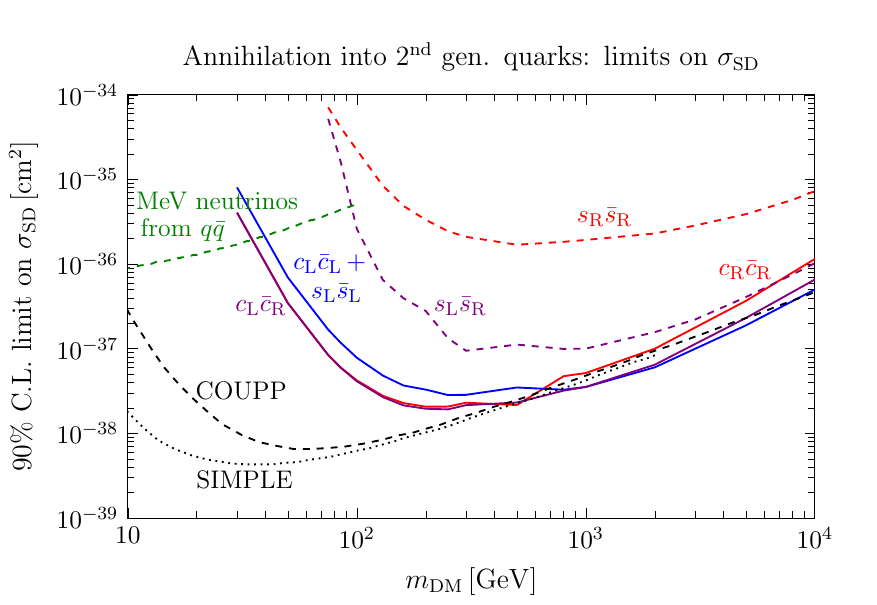}
\includegraphics[width=0.49\textwidth]{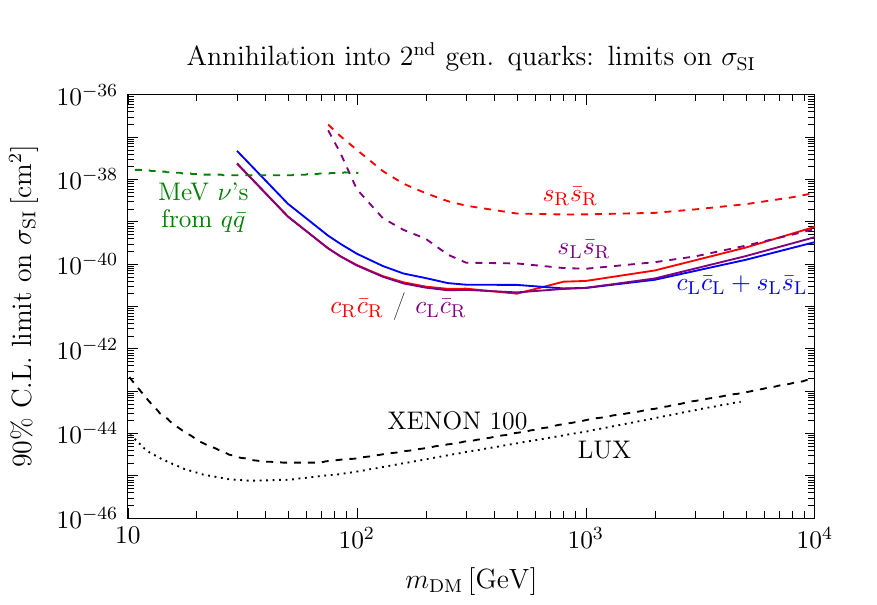} \\
\caption{\small 90\% C.L. limits on the spin-dependent (left plots) and spin-independent (right plots) interaction cross section from the non-observation of a high energy neutrino flux in the direction of the Sun in scenarios where the dark matter particle annihilates via a contact interaction to first and second generation leptons (top plots), first generation quarks (middle plots) or second generation quarks (lower plots). The green dashed line shows the limit derived in \cite{Bernal:2012qh} from the MeV neutrino flux produced in the annihilation into a fermion-antifermion pair, while the red, blue and purple lines are the limits derived in this paper from considering the final state radiation of gauge bosons off the external legs. We also show for comparison the best limits on the scattering cross section from direct detection experiments.
} 
\label{fig:contact-FSR}
\end{center}
\end{figure}

The limits on the  spin-dependent and spin-independent interaction cross sections are shown, respectively, in the left and right plots of \Figref{fig:contact-FSR}, for the annihilations into first and second generation leptons (top plots), first generation quarks (middle plots) or second generation quarks (lower plots). We also show for comparison the limits from the direct search experiments  COUPP \cite{Behnke:2012ys} and SIMPLE \cite{Felizardo:2011uw}, for the spin-dependent limits, or XENON100~\cite{Aprile:2012nq} and LUX~\cite{Akerib:2013tjd}, for the spin-independent limits, as well as the limits reported in \cite{Bernal:2012qh} from considering just the two-to-two annihilation into a fermion-antifermion pair and which only produces MeV neutrinos. We find that, for scenarios where the dark matter particle couples just to light fermions, the inclusion of the final state radiation allows to probe regions of the parameter space which were unconstrained by previous searches for dark matter annihilations in the Sun. 
Remarkably, the limits on the spin-dependent cross section derived in this paper are, in some instances, comparable to the limits from COUPP and SIMPLE; the limits from XENON100 and LUX are, as expected, much stronger than the spin-independent limits from IceCube.

As apparent from the plots, the limits on the cross section for annihilations into $f_R \bar f_R$, with $f=e,u,d$, are all comparable. In the leptonic case the only source of high energy neutrinos is the decay of the $Z$ boson produced in the electroweak bremsstrahlung process ${\rm DM}\,{\rm DM}\rightarrow f_R\bar f_R Z$, while in the hadronic case also the gluon bremsstrahlung ${\rm DM}\,{\rm DM}\rightarrow f_R\bar f_R g$ is relevant. The neutrino flux originated in the final state $f_R\bar f_R Z$ is proportional to the hypercharge of the fermion squared. Besides, we numerically find the contribution from the gluon bremsstrahlung to the total neutrino flux comparable to the contribution from the electroweak bremsstrahlung. As a result, the difference among the limits for $f=e,u,d$ is of ${\cal O}(1)$. We also show in the plot the limits for the final states $\mu_R \bar \mu_R$ and $s_R \bar s_R$. 
Since fermions of different generations have identical gauge quantum numbers, the limits for these two final states are identical to those for $e_R \bar e_R$ and $d_R \bar d_R$, respectively. Lastly, the limits for annihilations into $c_R\bar c_R$ are much stronger than in the other channels, due to the production of heavy hadrons, and in turn energetic neutrinos, in the hadronization of the charm (anti-) quarks.

On the other hand, the limits for the annihilations $f_L \bar f_R$ significantly differ depending on whether $f$ is a charged lepton or a quark.\footnote{Here and in the following, $f_L \bar f_R$ denotes the annihilation $f_L \bar f_R + f_R \bar f_L$, i.e. we assume $CP$ invariance in the annihilation process.} In this scenario, apart from the emission of a soft $Z$ boson off the fermions in the final state, it is also possible the emission of a soft $W$ boson off the left-handed fermion. In the case of the lepton, this process produces a hard neutrino with an energy close to the dark matter mass, while the neutrinos produced in the decay of the soft weak gauge boson from the final state radiation have much lower energies. 
As a result, the limits on the channel  $e_L \bar e_R$/$\mu_L \bar \mu_R$ are significantly stronger than for $u_L \bar u_R$ or $d_L \bar d_R$/$s_L \bar s_R$. Note also that the limits for the channels $q_L \bar q_R$ are stronger than for $q_R \bar q_R$, which is due to the hypercharge assignments of the quarks in the final state and due to existence of one additional annihilation channel, $q_L \bar q_R W$.

Lastly, for annihilations into $f_L \bar f_L$ again the limits for the leptonic channels are much stronger than for the hadronic channels. Due to the $SU(2)_L$ invariance, the annihilation necessarily produces both particles of the doublet with comparable rates, since the scale of the contact interaction (related to the mass of the new particles that induce the annihilation) is assumed to be much larger than the scale of electroweak symmetry breaking. More specifically, the possible final states including light left-handed leptons or quarks are $e_L \bar e_L+\nu_e \bar \nu_e$,  $\mu_L \bar \mu_L+\nu_\mu \bar \nu_\mu$, $u_L\bar u_L+d_L\bar d_L$ and $c_L\bar c_L+s_L\bar s_L$. In the first two cases, the two-to-two annihilation already produces hard neutrinos, as well as in the fourth case, from the hadronization of the charm quark. 
Hence, the limits on the cross section in this scenario are expected to be rather stringent for couplings to first and second generation left-handed leptons, as can be seen from the plot, and to a lesser extent, for couplings to second generation left-handed quarks.  In contrast, for annihilations into first generation left-handed quarks,  the emission of $W$ and $Z$ bosons off the final state is necessary in order to produce hard neutrinos, resulting into weaker limits.

\subsection{Loop annihilations dominant}
\label{sec:loop-dominant}

In some models the lowest order annihilation process into a fermion-antifermion pair can be s-wave and p-wave suppressed. A notable example are scenarios where the dark matter particle is a Majorana fermion that couples to a a light fermion of a definite chirality. In these scenarios, the s-wave annihilation into $f_R \bar f_R$ (or $f_L \bar f_L$) is helicity suppressed, concretely by $m^2_f/m_{\rm DM}^2$, while the p-wave annihilation is suppressed by the small velocity of the dark matter particles inside the Sun. Higher order processes, however, might not be s-wave suppressed. In particular, the loop induced annihilation into two gauge bosons, despite being loop suppressed, is not helicity suppressed and might have a much larger rate than the two-to-two annihilation into a fermion-antifermion pair (or the related two-to-three annihilation with the associated emission of a gauge boson).

Depending on the gauge quantum numbers of the fermions involved in the annihilation, various final states are possible: $\gamma\gamma$, $\gamma Z$, $Z Z$, $W^+ W^-$ or $g g$. The annihilation cross sections depend on the concrete Lorentz structure of the contact interaction, however in the limit $m_{\rm DM}\gg m_{f}$ the cross sections of the different channels are in simple relations, which are shown in Table \ref{table:x-section-loop}. In this table, $C_{\text{loop}}$ is a common factor of all cross sections, which depends on the nature of the effective coupling and which factors out when calculating the branching fractions. Besides, $\Sigma_{\gamma Z}$, $\Sigma_{Z Z}$,  $\Sigma_{WW}$ are phase space suppression factors which depend on the concrete effective operator inducing the annihilation and which take different values depending on whether parity is conserved or not in the annihilation process. The phase space suppression factors  were derived in \cite{Chen:2013gya} for s-wave annihilation of Majorana fermions into gauge bosons and read:
\begin{align}
\Sigma_{\gamma Z} &= \left( 1-\frac{m_Z^2}{4 m_{\rm DM}^2} \right)^3 \quad \text{for all operators}\;,\\
\Sigma_{VV} &= \begin{doublecases} 
&
\left( 1-\frac{m_V^2}{m_{\rm DM}^2} \right)^{1/2} \left( 1-\frac{m_V^2}{m_{\rm DM}^2}+\frac{3 m_V^4}{8 m_{\rm DM}^4} \right) \quad &\text{for }\bar{\chi} i \gamma^5 \chi B_{\mu \nu} B^{\mu \nu} \text{ or } \, \bar{\chi} i \gamma^5 \chi W^a_{\mu \nu} W^{a \mu \nu}\;,\\ 
&\left( 1-\frac{m_V^2}{m_{\rm DM}^2} \right)^{3/2} \quad &\text{for }\bar{\chi} i \gamma^5 \chi B_{\mu \nu} \widetilde{B}^{\mu \nu} \text{ or } \, \bar{\chi} i \gamma^5 \chi W^a_{\mu \nu} \widetilde{W}^{a \mu \nu}\;,
\end{doublecases}
\label{eq:Kinematic-suppression}
\end{align}
where in the last equation $V$ can be either a $Z$ or a $W$ boson.

\begin{table}
\begin{center}
\begin{tabular}{|c||c|c|c|c|}
\hline 
&
$\left( \sigma v \right)$ \\
\hline \hline
$\gamma\gamma$ & $N_C^2 \, \alpha_{\text{em}}^2 \, \left( 2I+1\right)^2 \, \left( I^2+Y^2\right)^2 \cdot C_{\text{loop}}$\\
\hline
$\gamma Z$ & $2 N_C^2 \, \alpha_{\text{em}}^2 \, \left[ \left(2I+1\right)Y^2\, \text{tan} \left( \theta_{\text{W}}\right)-2 I^2\, \text{cot} \left( \theta_{\text{W}}\right) \right]^2 \cdot C_{\text{loop}} \cdot \Sigma_{\gamma Z}$\\
\hline
$Z Z$ & $N_C^2 \, \alpha_{\text{em}}^2 \, \left[ \left(2I+1\right)Y^2\, \text{tan}^2 \left( \theta_{\text{W}}\right)+2 I^2\, \text{cot}^2 \left( \theta_{\text{W}}\right) \right]^2 \cdot C_{\text{loop}} \cdot \Sigma_{Z Z}$\\
\hline
$W^+ W^-$ & $2 N_C^2 \, \alpha_{\text{em}}^2 \, \left[ I^2/\text{sin}^4 \left( \theta_{\text{W}} \right) \right] \cdot C_{\text{loop}} \cdot \Sigma_{WW}$\\
\hline
$gg$ & $\left( N_C - 1 \right) \, \alpha_S^2 \, \left( 2I+1\right)^2 \cdot C_{\text{loop}}$\\
\hline
\end{tabular}
\end{center}
\caption{\small Annihilation cross section of the various loop annihilation channels into two gauge bosons in scenarios where the dark matter particle couples to a fermion-antifermion pair via a contact interaction. Here, $N_C=3$ for quarks and $N_C=1$ for leptons, $I=1/2$ for coupling to $SU(2)_L$ doublets and $I=0$ for singlets, and $Y$ is the hypercharge of the corresponding Standard Model fermion.}
\label{table:x-section-loop}
\end{table}

The upper limit on the scattering cross section in this scenario, $\sigma_{\rm SD/SI}^{\rm max}$, can be approximately calculated from the branching fractions in the various channels, ${\rm BF}_i$, and the limits on the scattering cross section in the corresponding channel $\sigma_{\rm SD/SI}^{{\rm max},i}$ (calculated assuming ${\rm BF}_i=1$):\footnote{ Due to the different choices of the opening cone angle in determining the limits in each of the channels, the upper bound obtained from the approximate expression \Equref{eq:CombiningLimits} differs from the upper bound directly calculated from the total flux. All results shown in this work are obtained using the full numerical approach; the difference with respect to \Equref{eq:CombiningLimits} is always less than 30\%.}
\begin{align}
 \frac{1}{\sigma_{\rm SD/SI}^{\rm max}(m_{\rm DM})} \simeq \sum\limits_{i}{\frac{{\rm BF}_i}{\sigma_{\rm SD/SI}^{{\rm max},i} (m_{\rm DM})}} \,.
 \label{eq:CombiningLimits}
\end{align}
The values of $\sigma_{\rm SD/SI}^{{\rm max}, W^+W^-}(m_{\rm DM})$ are equal to $\sigma_{\rm SD/SI}^{{\rm max}, ZZ}(m_{\rm DM})$ to a $10 \%$ accuracy. Besides, the upper limit on the scattering cross section for the $\gamma Z$ channel can be calculated from the upper limit for the $ZZ$ channel by weighting by the following factor:
\begin{align}
\sigma_{\rm SD/SI}^{{\rm max}, \gamma Z} \left(m_{\rm DM}\right) = \frac{2 \, \Gamma_C \left( m_{\rm DM}+\frac{m_Z^2}{4 m_{\rm DM}} \right)}{\Gamma_C \left( m_{\rm DM} \right)} \sigma_{\rm SD/SI}^{{\rm max}, ZZ} \left( m_{\rm DM}+\frac{m_Z^2}{4 m_{\rm DM}} \right)\;,
\end{align}
where $\Gamma_C(m_{\rm DM})$ is given in \Equref{eq:CaptureRate}. This relation comes from the fact that the energy of the $Z$ boson, and therefore the neutrino energy spectrum, produced in the annihilation of dark matter particles with mass $m_{\rm DM}$ into $\gamma Z$ is identical to the energy of the $Z$ boson produced in the annihilation of dark matter particles with mass $m_{\rm DM}+\frac{m_Z^2}{4m_{\rm DM}}$ into $ZZ$, and with a multiplicity which is in the former case a factor of two smaller than in the latter case. Therefore, \Equref{eq:CombiningLimits} can be cast as:
\begin{align}
 \frac{1}{\sigma_{\rm SD/SI}^{\rm max}\left(m_{\rm DM}\right)}  = 
 &\frac{{\rm BF}_{ZZ}+{\rm BF}_{W^+W^-}}{\sigma_{\rm SD/SI}^{{\rm max}, ZZ} \left(m_{\rm DM}\right)} + 
\frac{1}{2}  \frac{\Gamma_C\left(m_{\rm DM}\right)}{\Gamma_C  \left(m_{\rm DM} + \frac{m_Z^2}{4m_{\rm DM}}\right)} \frac{{\rm BF}_{\gamma Z}}{\sigma_{\rm SD/SI}^{{\rm max}, ZZ} \left(m_{\rm DM} + \frac{m_Z^2}{4m_{\rm DM}}\right)}
 \nonumber \\
&+ \frac{{\rm BF}_{gg}}{\sigma_{\rm SD/SI}^{{\rm max}, gg} \left(m_{\rm DM}\right)} \,,
 \label{eq:CombiningLimitsConcrete}
\end{align}
which just depends on the branching fractions of the various channels, which can be calculated from the values of $\left( \sigma v \right)$ given in Table \ref{table:x-section-loop}, as well as on the upper limits on the scattering cross section in the channels $ZZ$ and $gg$; the values of $\sigma_{\rm SD/SI}^{{\rm max}, gg}$ and $\sigma_{\rm SD/SI}^{{\rm max}, ZZ}$ relevant for the evaluation of \Equref{eq:CombiningLimitsConcrete} are reported in Appendix \ref{ap:ZZ-gg} for various values of the dark matter mass.

\begin{figure}[h!]
\begin{center}
\includegraphics[width=0.49\textwidth]{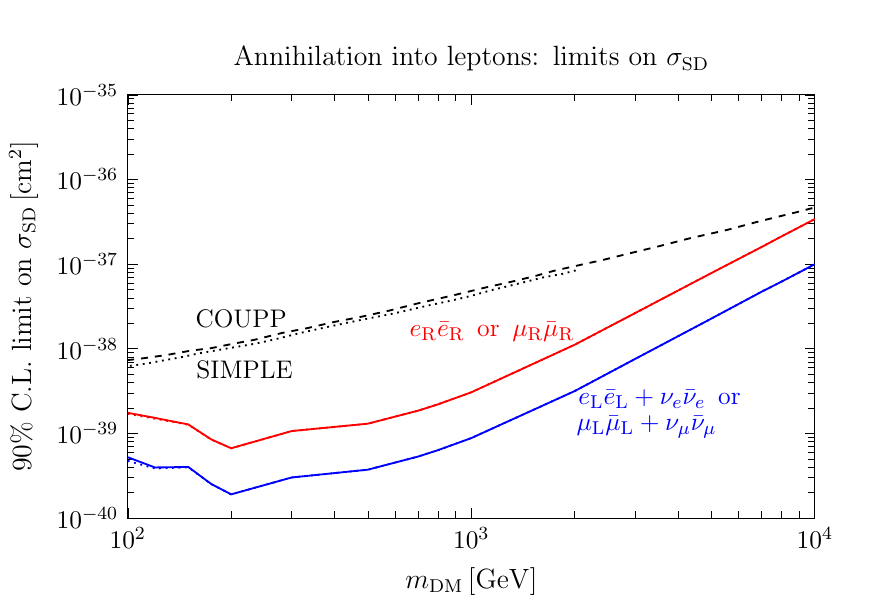}
\includegraphics[width=0.49\textwidth]{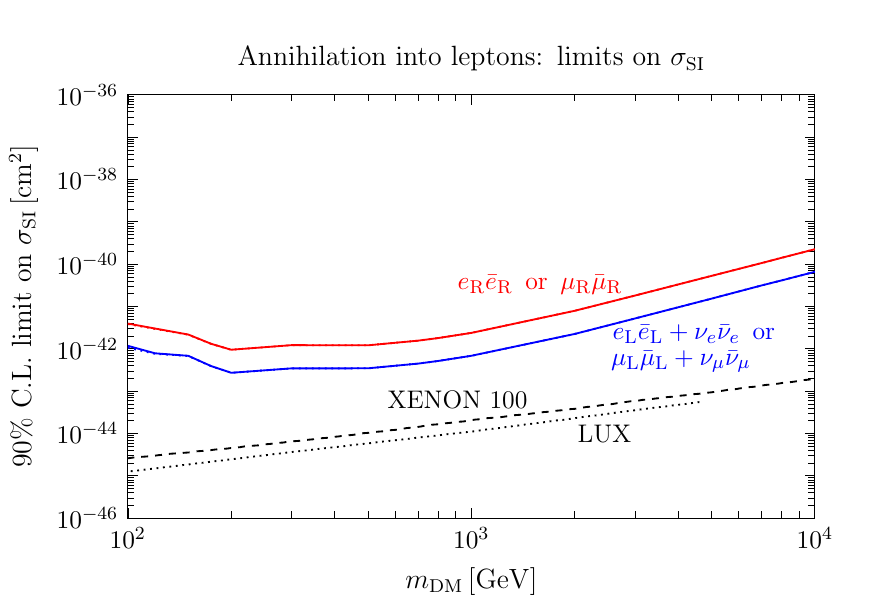} \\
\includegraphics[width=0.49\textwidth]{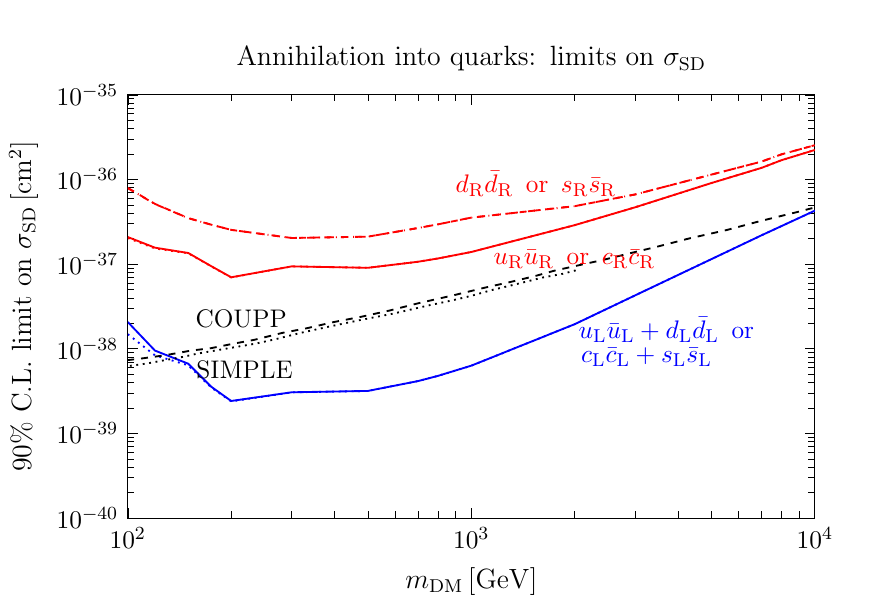}
\includegraphics[width=0.49\textwidth]{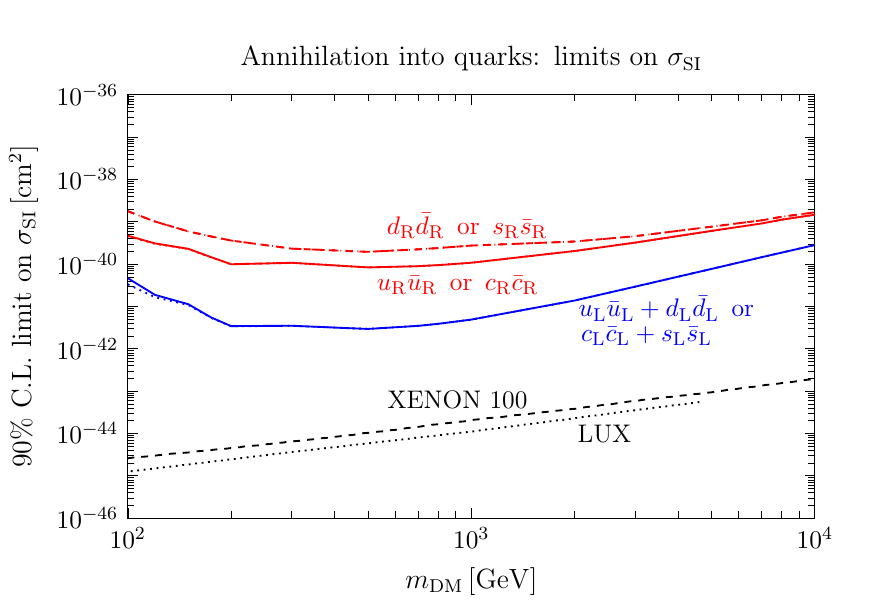} \\
\caption{\small 90\% C.L. limits on the spin-dependent (left plots) and spin-independent (right plots) interaction cross section from the non-observation of the high-energy neutrino flux produced in the one loop annihilation into two gauge bosons in scenarios where the dark matter particle couples to leptons (top plots) or to quarks (bottom plots) via a contact interaction. We also show for comparison the best limits on the scattering cross section from direct detection experiments.} 
\label{fig:contact-Loops}
\end{center}
\end{figure}

We show in \Figref{fig:contact-Loops} the resulting limits on the spin-dependent (left plots) and spin-independent (right plots) scattering cross section with protons for loop-dominated annihilations arising in scenarios where the dark matter particle couples to leptons (upper plots) or to quarks (lower plots), together with the best limits from direct search experiments. In the plot we show the results for the two possible phase space suppression factors for annihilations into two gauge bosons, depending on whether the contact interaction is $P$-even or $P$-odd, in \Equref{eq:Kinematic-suppression}. Remarkably, the limits are quite insensitive to this model dependent factor, even close to the production threshold. 

It is apparent from the plot that the limits for dark matter scenarios with couplings to leptons are more stringent than for couplings to quarks, due to the larger branching fraction of the channels with weak gauge bosons in the final state, which produce hard neutrinos. It is also apparent from the plot that the limits for dark matter scenarios with couplings to left-handed particles are stronger than for couplings to right-handed particles, which is due to the additional annihilation channel $W^+ W^-$ in the former case as well as the different hypercharges of the fermions circulating in the loop. 
It is interesting that the IceCube limits on the spin-dependent interaction cross section are, for the leptonic case, significantly stronger than the limits from direct search experiments; for the hadronic case the IceCube limits are slightly weaker than the direct detection limits for couplings to right-handed quarks and slightly stronger for couplings to left-handed quarks. In contrast, the limits for the spin-independent interaction cross section from the direct search experiments XENON100 and LUX are stronger than the IceCube limits.

\section{Toy model with Majorana fermions as dark matter}
\label{sec:toy-model}

In a model where the dark matter particle couples to fermions, the two-to-two annihilations into a fermion-antifermion pair, the two-to-three annihilations with the associated emission of a gauge boson and the one loop annihilation into two gauge bosons will necessarily occur. However, the branching fraction for each of these processes depends on the details of the model, and accordingly the IceCube limits on the model parameters. 

In this section we discuss in detail a well motivated class of dark matter models where the dominant annihilation channels are the higher order two-to-three and loop processes, due to the helicity and velocity suppression of the two-to-two annihilation cross section into light fermions. Concretely, we will analyze a toy model where the dark matter particle is a Majorana fermion $\chi$, singlet under the Standard Model gauge group, that interacts with a light right-handed fermion $f_R$ and a scalar $\eta$ via a Yukawa interaction with coupling strength $y$. The Lagrangian of the model is given by
\begin{align}
  {\cal L}={\cal L}_{\rm SM}+{\cal L}_{\chi}+{\cal L}_\eta+ {\cal L}_{\rm int}
  \;.
\end{align} 
Here, ${\cal L}_{\rm SM}$ is the Standard Model Lagrangian, while ${\cal L}_{\chi}$ and ${\cal L}_\eta$ are the parts of the Lagrangian involving just the new fields $\chi$ and $\eta$ and which read, respectively
\begin{align}
 \begin{split}
    {\cal L}_\chi&=\frac12 \bar \chi^c i\slashed {\partial} \chi
    -\frac{1}{2}m_\chi \bar \chi^c\chi\; \; \text{and}\\ {\cal L}_\eta&=(D_\mu
    \eta)^\dagger  (D^\mu \eta)-m_\eta^2 \eta^\dagger\eta -\frac{1}{2}\lambda_2 (\eta^\dagger \eta)^2\;,
  \end{split}
\end{align}
where $D_\mu$ denotes the covariant derivative. Lastly, the interaction term in the Lagrangian is given by
\begin{align}
  {\cal L}_{\rm int} &= -\lambda_3(\Phi^\dagger \Phi)(\eta^\dagger \eta) - y \bar \chi f_R \eta+{\rm h.c.} \;.
\label{eq:singlet-qR}
\end{align}
Here, $\Phi$ is the Standard Model Higgs doublet and $\eta$ is a scalar field with quantum numbers under $SU(3)_C\times SU(2)_L\times U(1)_Y$ which are $(1,1,1)$ for couplings to right-handed leptons while $(\bar{3},1,\frac{1}{3})$ and $(\bar{3},1,-\frac{2}{3})$ for couplings to down- and up-type right-handed quarks, respectively. Notice that all the masses and couplings of the dark sector can be taken real, hence CP is conserved. In what follows we will set for simplicity $\lambda_3=0$.

\begin{figure}[t]
  \centering
  \includegraphics[width=0.15\textwidth]{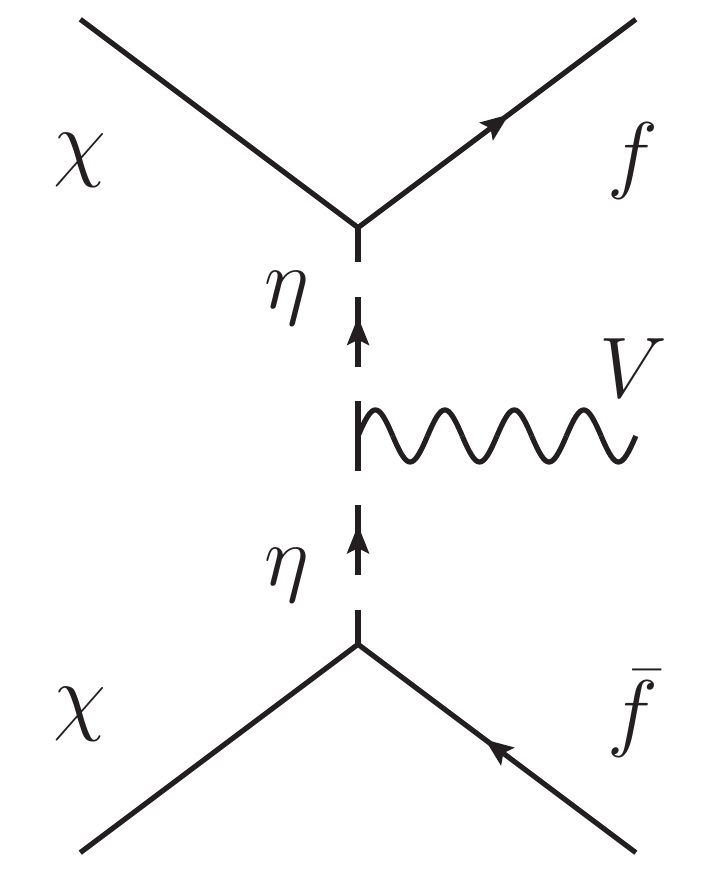}\hspace{1cm}
  \includegraphics[width=0.15\textwidth]{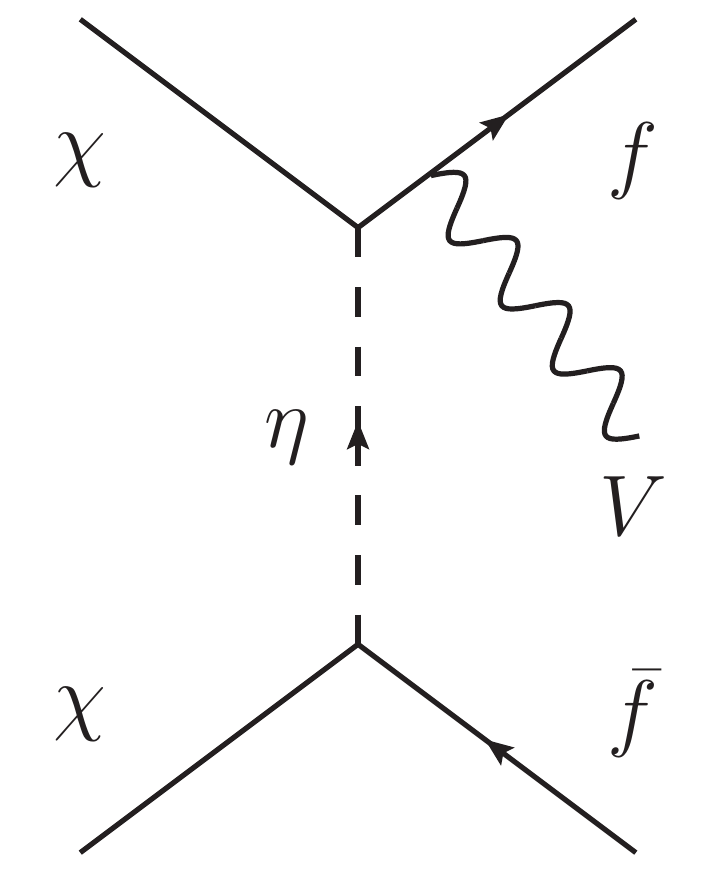}\hspace{1cm}
  \includegraphics[width=0.15\textwidth]{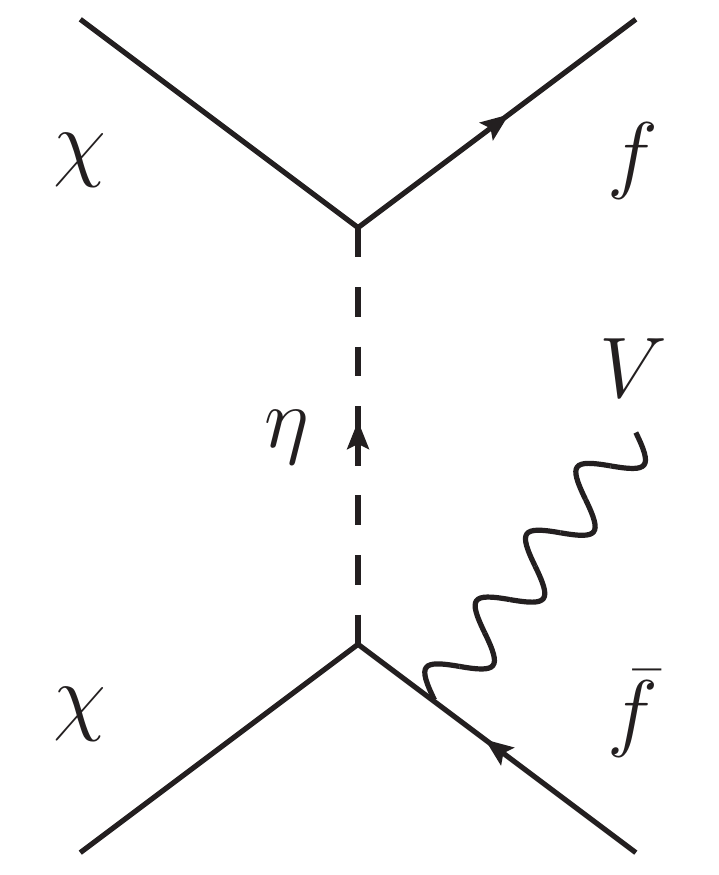}\\ \vspace{0.6cm}
  \includegraphics[width=0.2\textwidth]{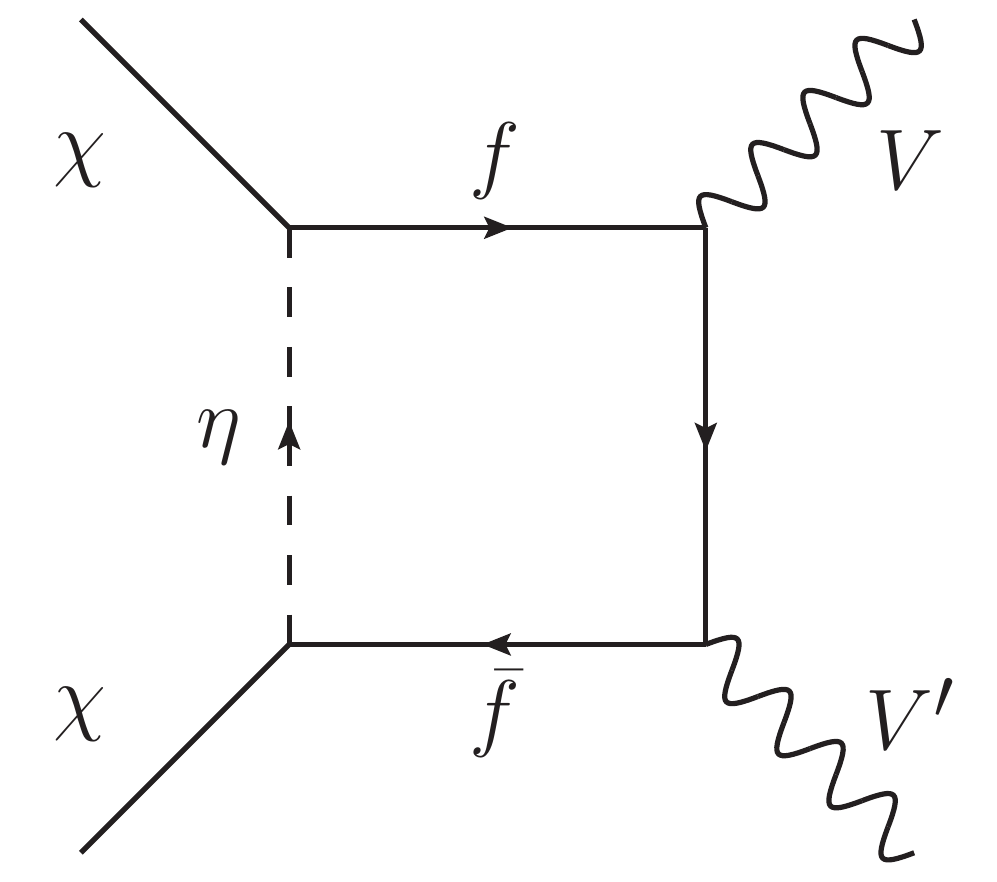}\hspace{1cm}
  \includegraphics[width=0.2\textwidth]{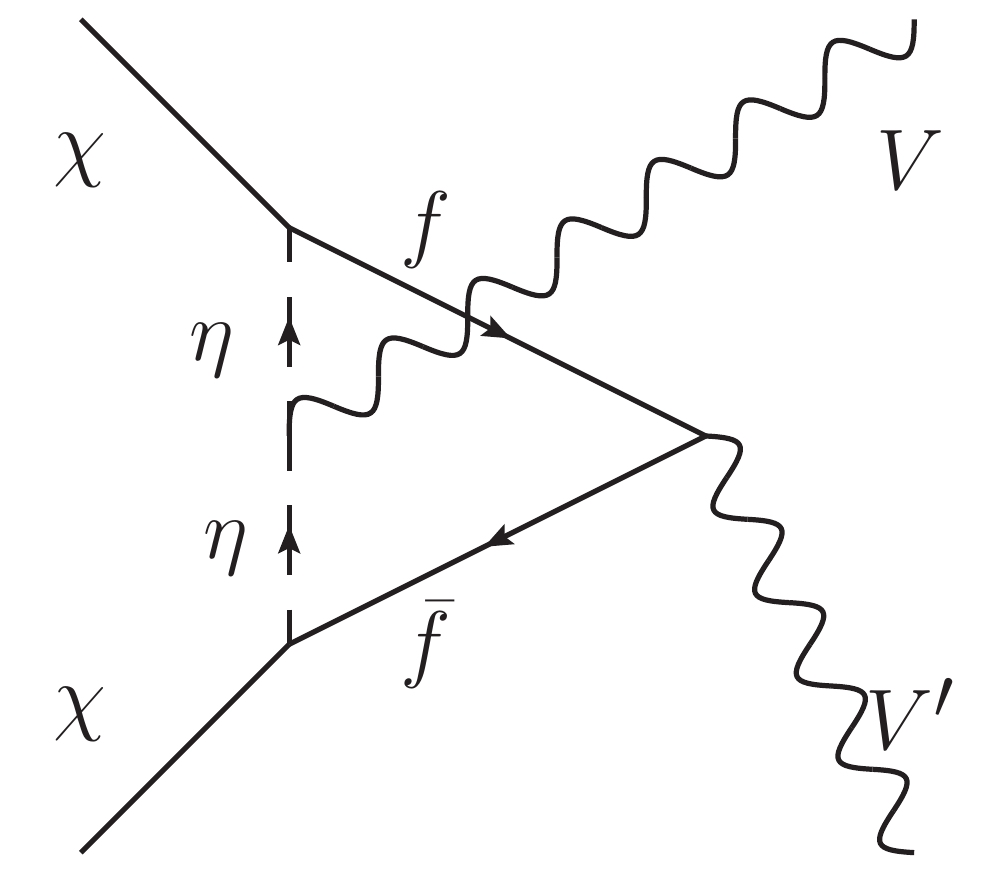}
  \caption{\small Feynman diagrams inducing the two-to-three annihilation with one gauge boson in the final state (top) and the annihilation into two gauge bosons at the one loop level (bottom).}
  \label{fig:toymodel-diagrams}
\end{figure}

The Feynman diagrams that induce the two-to-three and the loop processes in this toy model are shown in \Figref{fig:toymodel-diagrams} and the corresponding expressions for the cross sections are presented in Appendix~\ref{ap:crosssections}. For couplings to right-handed electrons or right-handed up/down quarks we find that the total annihilation cross section can be approximated by:

\begin{align}
 \left( \sigma_{\mathrm{ann}} v \right) \approx
 \begin{cases}
\displaystyle{\left(\frac{37.6}{\left(m_\eta/m_\chi\right)^8} + \frac{0.62}{\left(m_\eta/m_\chi\right)^4} \right) y^4 \left(\frac{\mathrm{TeV}}{m_\chi}\right)^2\times 10^{-30} \, \mathrm{cm^3s^{-1}}} \;,\; \mathrm{for~couplings~to~}e_R,  \vspace{0.3cm} \\
 \displaystyle{\left(\frac{1.2 \times 10^3}{\left(m_\eta/m_\chi\right)^8} + \frac{74.4}{\left(m_\eta/m_\chi\right)^4} \right) y^4 \left(\frac{\mathrm{TeV}}{m_\chi}\right)^2\times 10^{-30} \, \mathrm{cm^3s^{-1}}} \;,\; \mathrm{for~couplings~to~}u_R, \vspace{0.3cm} \\
 \displaystyle{\left(\frac{1.2 \times 10^3}{\left(m_\eta/m_\chi\right)^8} + \frac{73.3}{\left(m_\eta/m_\chi\right)^4} \right) y^4 \left(\frac{\mathrm{TeV}}{m_\chi}\right)^2\times 10^{-30} \, \mathrm{cm^3s^{-1}}} \;,\; \mathrm{for~couplings~to~}d_R,
 \end{cases}
 \label{eq:AnnihilationCSOrderScaling}
\end{align}
which differ from the exact results by less than $\sim 40 \%$ throughout the range $100$ GeV $\leq m_\chi \leq 10$ TeV and $1.01 \leq m_\eta/m_\chi \leq 10$. In \Equref{eq:AnnihilationCSOrderScaling}, the terms scaling as $\left(m_\eta/m_\chi\right)^{-8}$ arise from the two-to-three processes, while those scaling as $\left(m_\eta/m_\chi\right)^{-4}$ correspond to the loop diagrams. It follows from  \Equref{eq:AnnihilationCSOrderScaling} that for small mass splittings, $m_\eta/m_\chi\simeq 1$, the dominant annihilation process is the two-to-three channel, while for large mass splittings,  $m_\eta/m_\chi\gg 1$, the loop processes dominate; the transition between both regimes occurs at $m_{\eta} / m_{\chi} \sim 3 \left(2 \right)$ in the case of coupling to right-handed electrons (right-handed up- or down-quarks).

\begin{figure}[t]
\begin{center}
\includegraphics[width=0.49\textwidth]{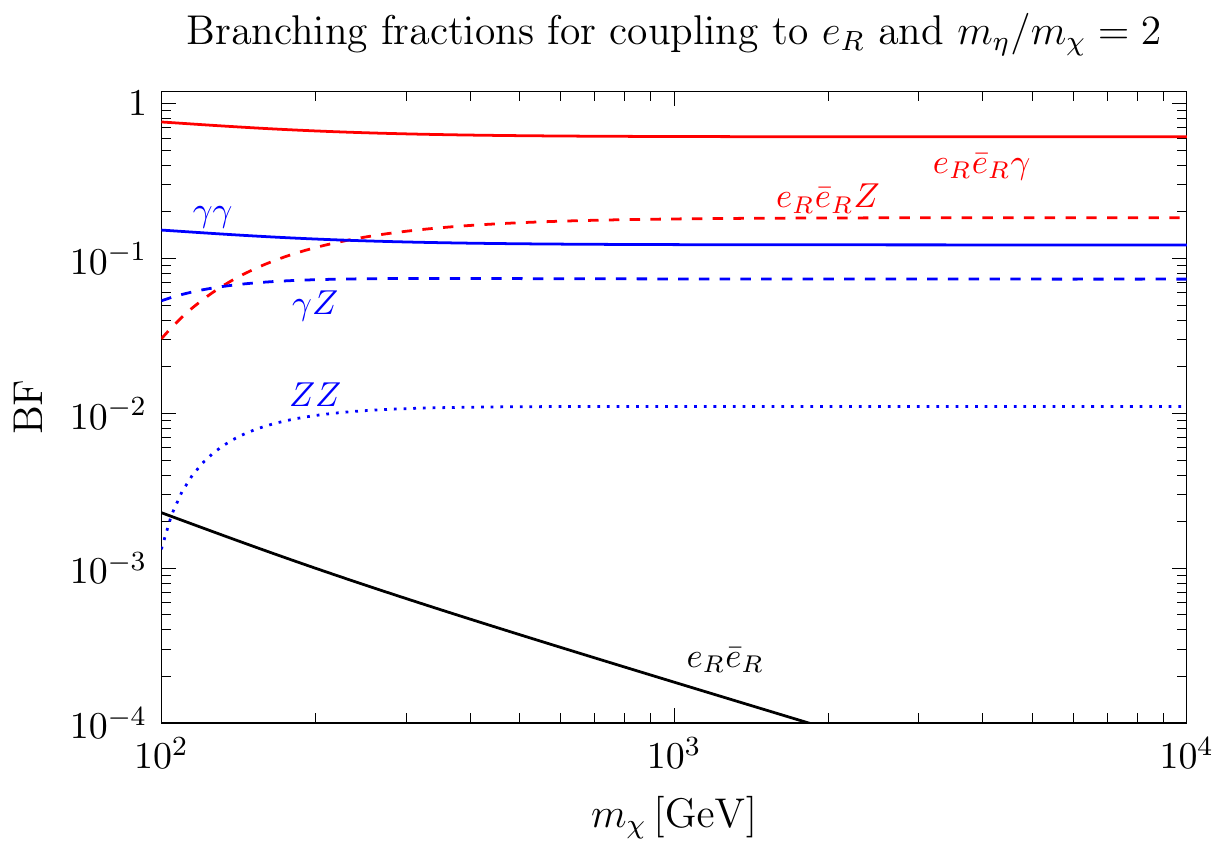}
\includegraphics[width=0.49\textwidth]{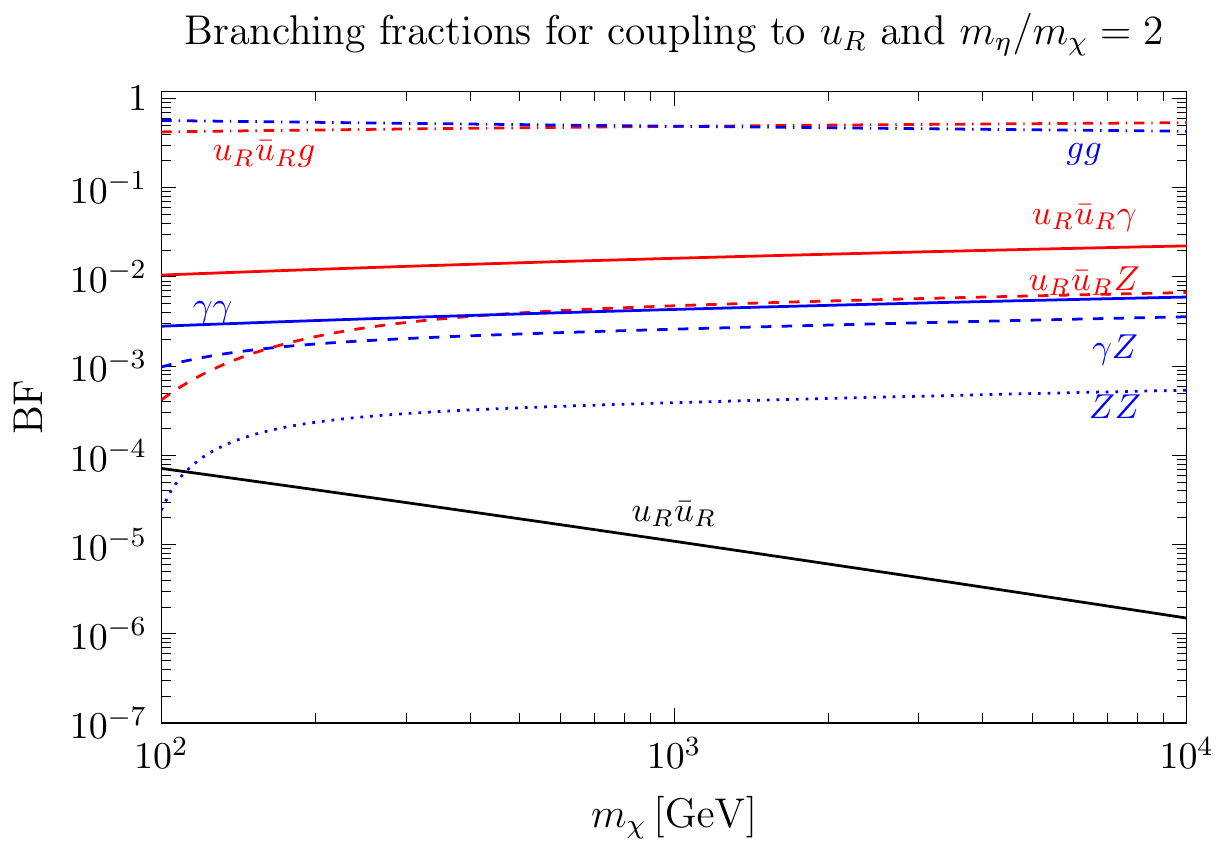} \\
\includegraphics[width=0.49\textwidth]{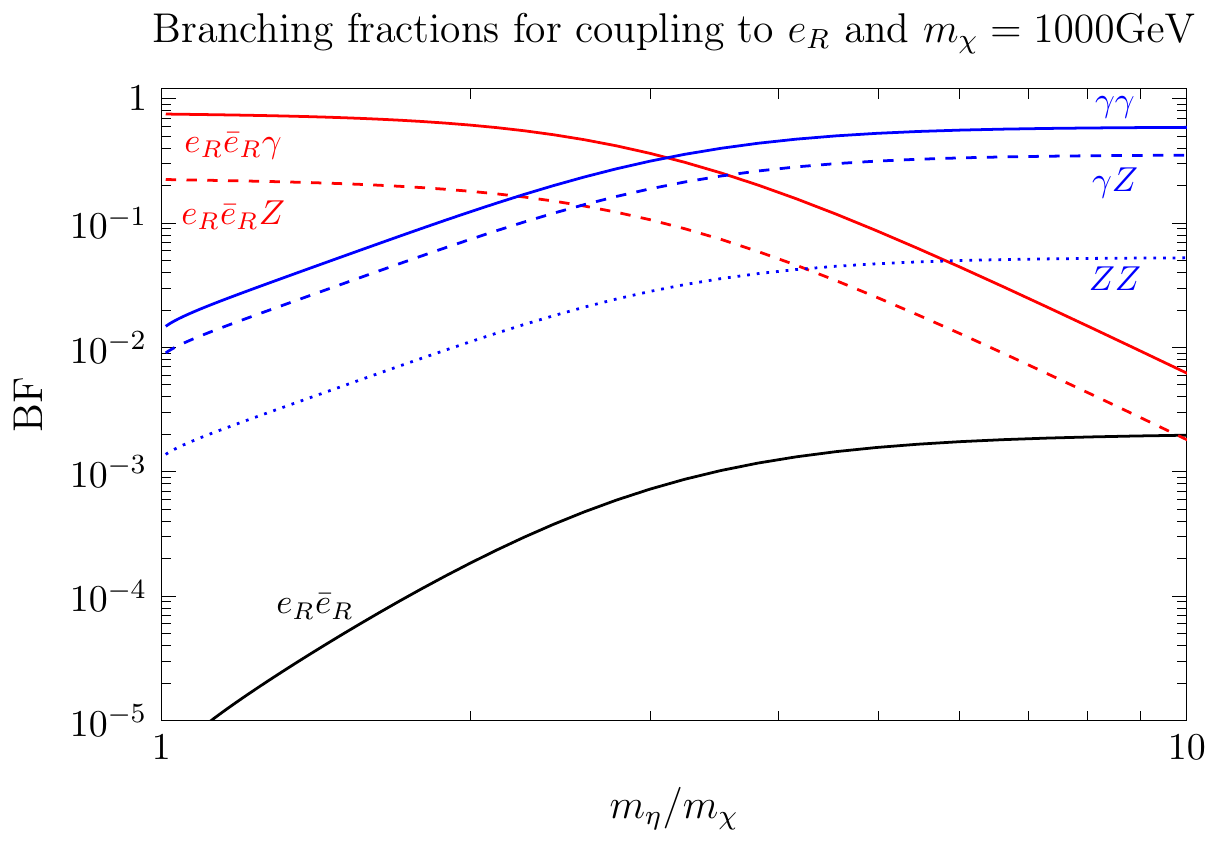}
\includegraphics[width=0.49\textwidth]{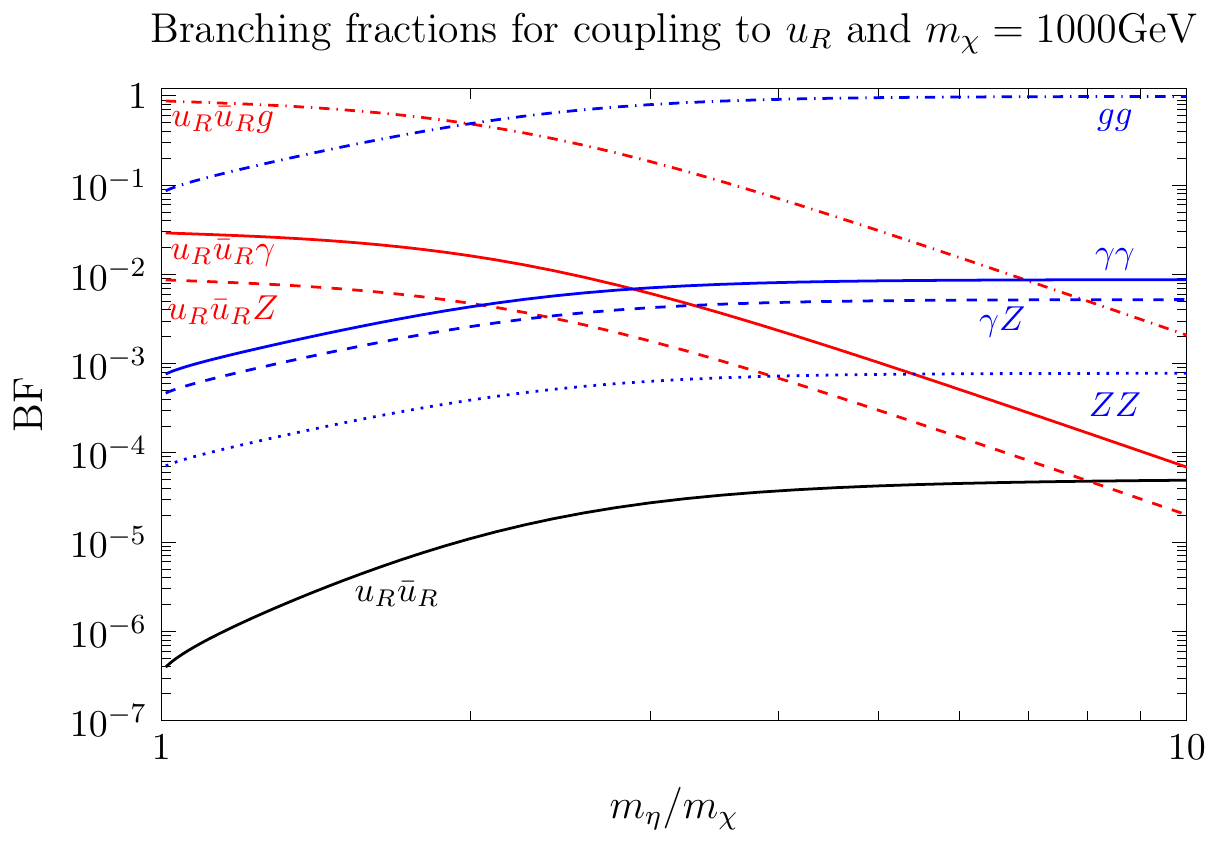} \\
\caption{\small Branching ratios of the various tree-level and one-loop annihilation channels for a toy model with a Majorana fermion as dark matter particle that couples to the electron (left plots) or to the up-quark (right plots) via a Yukawa coupling with a scalar. The top plots show the branching ratios as a function of the dark matter mass $m_\chi$ for fixed ratio of the scalar mass and the dark matter mass, $m_\eta/m_\chi$, while the bottom plots, for fixed $m_\chi=1000\GeV$ as a function of  $m_\eta/m_\chi$.}
\label{fig:BRs}
\end{center}
\end{figure}

For the analysis of the annihilation signals from the Sun, and under the assumption that dark matter captures and annihilations are in equilibrium in the solar interior, only the branching fractions are relevant, which depend on $m_\chi$ and $m_\eta$ or, alternatively, on $m_\chi$ and $m_\eta/m_\chi$. The branching fractions for couplings to right-handed electrons (right-handed up-quarks) are shown in the left plot (right plot) of \Figref{fig:BRs} for $m_\eta/m_\chi=2$ as a function of the dark matter mass (upper plots) and for $m_\chi=1000$ GeV as a function of $m_\eta/m_\chi$ (lower plots); the results for couplings to down-quarks are qualitatively similar to the case of the up-quarks.

\begin{figure}[t]
\begin{center}
\includegraphics[width=0.49\textwidth]{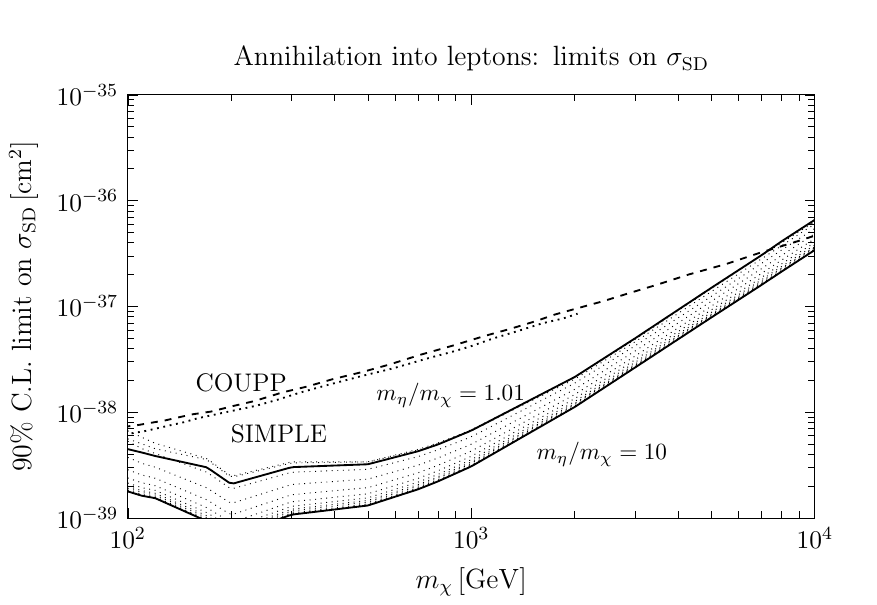}
\includegraphics[width=0.49\textwidth]{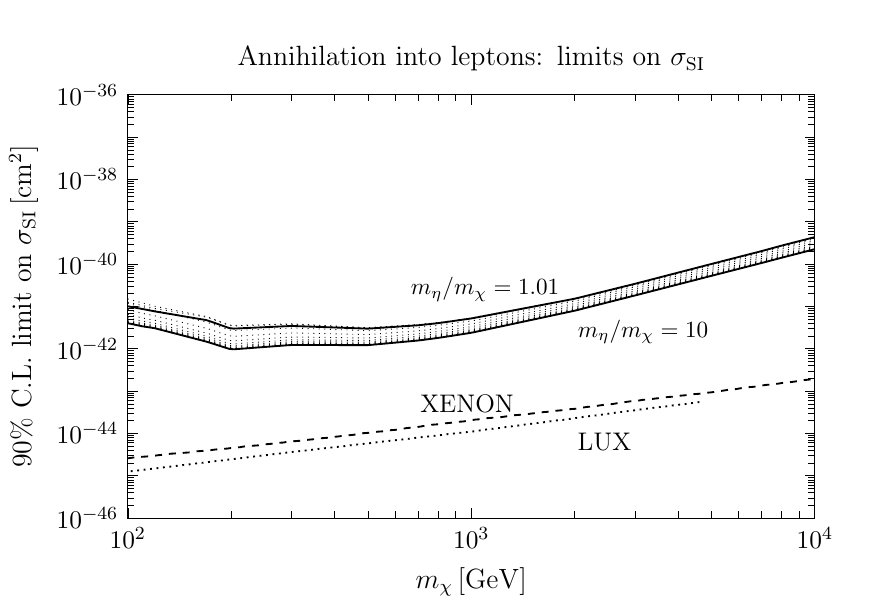} \\
\includegraphics[width=0.49\textwidth]{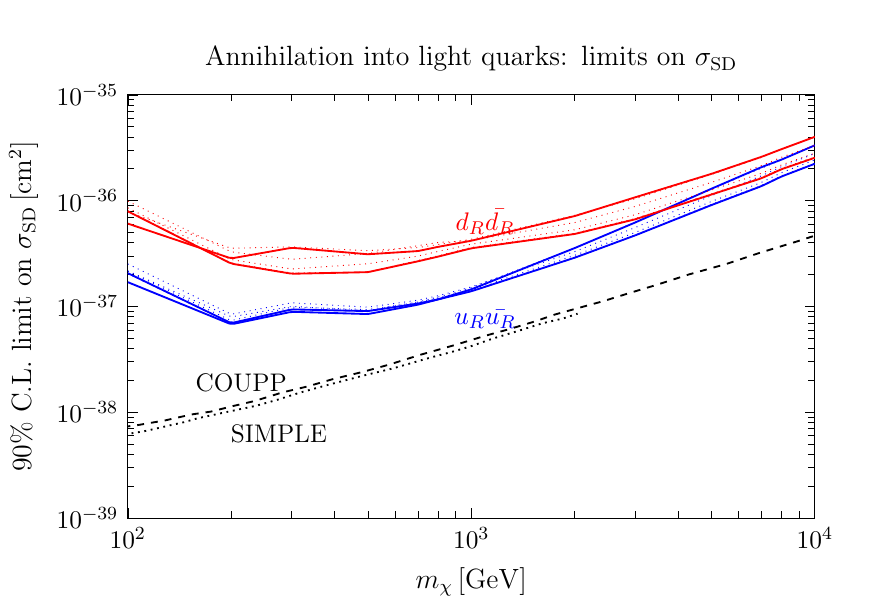}
\includegraphics[width=0.49\textwidth]{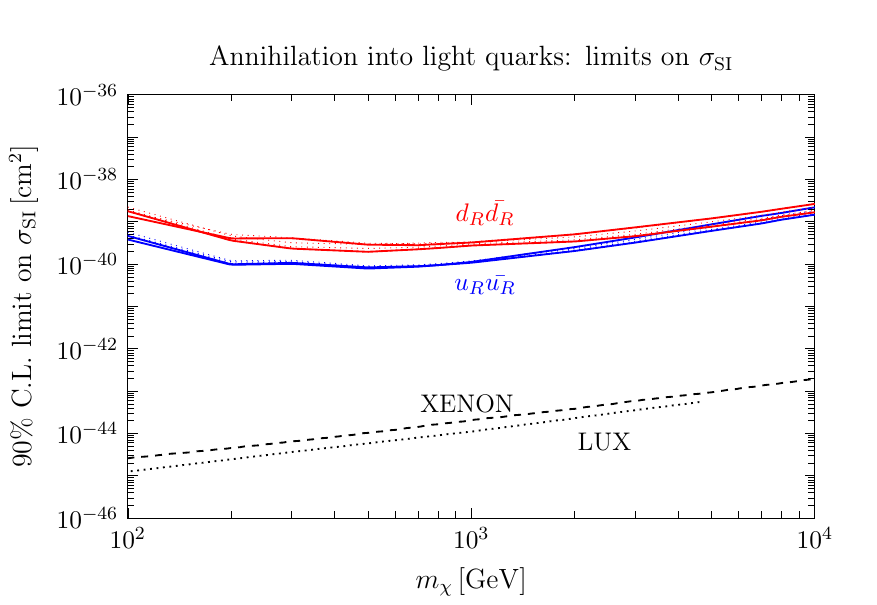} \\
\caption{\small 90\% C.L. limits on the spin-dependent (left plots) and spin-independent (right plots) interaction cross section as a function of the dark matter mass in a toy model with Majorana fermions as dark matter particles that couple to the right-handed electrons (top plots) or to the right-handed first generation quarks (lower plots), for various values of the mass ratio $m_\eta/m_\chi$. We also show for comparison the best limits on the scattering cross section from direct detection experiments. } 
\label{fig:limits-toy-model}
\end{center}
\end{figure}

The channels producing hard neutrinos are $f_R \bar f_R Z$ (dashed red line), $\gamma Z$ (dashed blue line), $ZZ$ (dotted blue line) and, in the case of couplings to quarks, $f_R \bar f_R g$ (dotted-dashed red line) and $gg$ (dotted-dashed blue line). As apparent from the plot, the branching fractions for these channels can be sizable in both the degenerate and the hierarchical scenario. Approximate expressions for the branching fractions in the relevant channels can be found in Table \ref{tab:BRsAnalytic}; the numerical values in these formulas have been obtained evaluating, for illustration, the strong coupling constant at the scale $m_{\chi}=1 \TeV$.

\def\tabularxcolumn#1{m{#1}}
\newcolumntype{C}[1]{>{\centering\arraybackslash}p{#1}}
{\renewcommand{\arraystretch}{0.2}
\begin{table}[htbp]
 \centering
 \begin{tabular}{ccC{2.9cm}C{0.3cm}C{1.5cm}cC{2.8cm}cc}
 \toprule
 & BF & \multicolumn{3}{c}{$f= e$} &\quad \quad& \multicolumn{3}{c}{$f=u$}
\\
 \midrule
 \multirow{5}{*}{\rotatebox{90}{degenerate~limit}}
&  ${\rm BF}_{f_R\bar f_R Z}$&$\sin^2\theta_{\mathrm{W}}$&$\simeq$&$0.23$    
&\quad \quad& $\frac{\alpha_{\mathrm{em}} \tan^2
\theta_{\mathrm{W}}}{3 \alpha_{\mathrm{s}}(m_\chi) }$&$\simeq$& $
10^{-2}$ \\[2ex]
  &    ${\rm BF}_{\gamma Z}$&$\frac{\pi^3
\alpha_{\mathrm{em}} \sin^2  \theta_{\mathrm{W}} }{32 \left(
7/2-\pi^2/3\right)}$&$\simeq$&$8 \cdot 10^{-3}$    &\quad \quad&$ \frac{\pi^3\alpha_{\mathrm{em}}^2
\tan^2\theta_{\mathrm{W}} }{72
\left( 7/2-\pi^2/3\right) \alpha_{\mathrm{s}}(m_\chi)}$&$ \simeq$&$5 \cdot
10^{-4}$\\[2ex]
                        &    ${\rm BF}_{Z
Z}$&$\frac{\pi^3\alpha_{\mathrm{em}} \sin^2
\theta_{\mathrm{W}} \tan^2
\theta_{\mathrm{W}} }{64 \left( 7/2-\pi^2/3\right)}$&$\simeq$&$10^{-3}$
    &\quad \quad&
$ \frac{\pi^3 \alpha_{\mathrm{em}}^2
\tan^4\theta_{\mathrm{W}} }{144 \left( 7/2-\pi^2/3\right)
\alpha_{\mathrm{s}}(m_\chi)}$&$ \simeq$&$ 7 \cdot 10^{-5}$ \\[2ex]
                   & ${\rm BF}_{gg}$    &    &--&          &\quad \quad&
                    $\frac{\pi^3 \alpha_{\mathrm{s}}(m_\chi)}{128 \left(
7/2-\pi^2/3\right)}$&$\simeq$&$ 9 \cdot 10^{-2}$ \\[2ex]
           &  ${\rm BF}_{f_R\bar f_R g}$             &    &--&         &\quad \quad& 
&$\simeq$&$ 1$ \\[2ex]
  \midrule
  \multirow{5}{*}{\rotatebox{90}{hierarchical~limit}}         &    ${\rm
BF}_{f_R\bar
f_R Z}$&&$\simeq$&$0$&\quad \quad&    &$\simeq$& $0$ \\[2ex]
                        &    ${\rm BF}_{\gamma Z}$&$\frac{2 \tan^2
\theta_{\mathrm{W}}}{(1+  \tan^2  \theta_{\mathrm{W}} )^2}$&$ \simeq$&$
0.38$
& \quad \quad&$\frac{16 \alpha_{\mathrm{em}}^2 \tan^2
\theta_{\mathrm{W}} }{9 \alpha_{\mathrm{s}}^2(m_\chi)}$&$ \simeq$&$ 5 \cdot
10^{-3}$\\[2ex]
                        &    ${\rm BF}_{Z Z}$&$\frac{\tan^4
\theta_{\mathrm{W}}}{(1+  \tan^2 \theta_{\mathrm{W}} )^2}$&$ \simeq$&$
5.6 \cdot
10^{-2}$     &\quad \quad& $ \frac{8
\alpha_{\mathrm{em}}^2 \tan^4
 \theta_{\mathrm{W}}}{9 \alpha_{\mathrm{s}}^2(m_\chi)}$&$ \simeq$&$ 8\cdot
10^{-4}$ \\
[2ex]
   &${\rm BF}_{gg}$                     &    &--&          &\quad \quad& &$ \simeq$&$
1$ \\[2ex]
  &${\rm BF}_{f_R\bar f_R
  g}$                      &    &--&         &\quad \quad& &$\simeq$&$ 0$ \\[2ex]
  \bottomrule
\end{tabular}
\caption{Branching fractions into the different channels in the degenerate limit (upper panel) and the hierarchical limit (lower panel). These formulae apply to the limit $m_\chi \gg m_Z$ where phase space effects are negligible. The numerical values were obtained evaluating the strong coupling constant at $1 \, \mathrm{TeV}$.}
\label{tab:BRsAnalytic}
\end{table}
}

The sizable branching fraction into these channels then allows to set constraints on this model from the non-observation of a significant neutrino excess in the direction of the Sun with respect to the expected atmospheric background. We calculate the neutrino spectra from the decay and hadronization of the gauge bosons using PYTHIA 8.176~\cite{Sjostrand:2006za,Sjostrand:2007gs} (interfaced with CalcHEP~\cite{Pukhov:1999gg,Pukhov:2004ca} in the case of the two-to-three processes) and we then derive limits on the interaction cross section with protons following the procedure described in Section~\ref{sec:loop-subdominant}. The resulting limits on the spin-dependent (left panel) and spin-independent (right panel) cross section are shown in \Figref{fig:limits-toy-model}, for dark matter annihilations mediated by couplings to right-handed electrons (upper plots) and right-handed up- or down-quarks (lower plots) as a function of the dark matter mass, for eight different values of the parameter $m_\eta/m_\chi$ ranging between $1.01$ and 10.

 It is noticeable from the plot that the limits are quite insensitive to the value of $m_\eta/m_\chi$. Namely between the degenerate case ($m_\eta/m_\chi\simeq 1$) and the hierarchical case ($m_\eta/m_\chi\gg 1$) the limits differ by approximately a mere factor of 2 for the case of couplings to $e_R$, between 1 and 1.5 in the case of couplings to $u_R$ and between 1 and 1.8 in the case of couplings to $d_R$. 
In the case of couplings to $e_R$ this result can be understood from the fact that in the degenerate case the most important annihilation channel producing hard neutrinos is the two-to-three annihilation into $e_R \bar e_R Z$, while in the hierarchical case it is the loop annihilation into $\gamma Z${. In the degenerate case, the energy of the $Z$-boson is close to the dark matter mass, therefore it is possible to derive an approximate limit on the spin-dependent and spin-independent interaction cross sections from the corresponding limits in the $ZZ$ channel given in  Appendix \ref{ap:ZZ-gg}. This limit reads
\begin{align}
 \sigma^{\rm max, deg}_{\rm SD/SI}\simeq \frac{2\sigma_{\rm SD/SI}^{{\rm max}, ZZ}}{{\rm BF}_{e_R\bar e_R Z}}\simeq   \frac{2\sigma_{\rm SD/SI}^{{\rm max},ZZ}}{\sin^2\left(\theta_W\right)}\;.
  \label{eq:LimitRescalingLeptonsDegenerate}
\end{align}
On the other hand, in the hierarchical case, since $m_{\eta}/m_{\chi} \gg 1$, the annihilation can be described by a contact interaction. Then, using Eq.(\ref{eq:CombiningLimitsConcrete}) one finds, in the limit $m_\chi\gg m_Z$,
\begin{align}
 \sigma^{\rm max, hier}_{\rm SD/SI}\simeq 
  \frac{\sigma^{{\rm max}, ZZ}_{\rm SD/SI}}{{\rm BF}_{ZZ}+{\rm BF}_{\gamma Z}/2}=
  \frac{\sigma^{{\rm max}, ZZ}_{\rm SD/SI}}{\sin^2\left(\theta_W\right)}\;,
 \label{eq:LimitRescalingLeptonsHierarchical}
\end{align}
which differs from \Equref{eq:LimitRescalingLeptonsDegenerate} by a factor of 2.

In the case of couplings to $u_R$ or $d_R$, channels involving a $Z$ boson as well as channels with a gluon in the final state also contribute to the limits. Concretely, in the case of couplings to $u_R$, a similar argument as above leads to
\begin{align}
 \sigma^{\rm max, deg}_{\rm SD/SI}\simeq \frac{2 \sigma_{\rm SD/SI}^{{\rm max}, gg}}{1+{\rm BF}_{u_R \bar{u}_R Z} \cdot \xi_{\rm SD/SI} \left(m_\chi \right)} \simeq \frac{2 \sigma_{\rm SD/SI}^{{\rm max}, gg}}{1+10^{-2} \cdot \xi_{\rm SD/SI} \left(m_\chi \right)}\;,
\label{eq:sigma_deg_uR}
\end{align}
and 
\begin{align}
 \sigma^{\rm max, hier}_{\rm SD/SI}\simeq \frac{ \sigma_{\rm SD/SI}^{{\rm max}, gg}}{1+\left({\rm BF}_{ZZ}+{\rm BF}_{\gamma Z}/2 \right) \cdot \xi_{\rm SD/SI} \left(m_\chi \right)} \simeq \frac{\sigma_{\rm SD/SI}^{{\rm max}, gg}}{1+3.5 \cdot 10^{-3} \cdot \xi_{\rm SD/SI} \left(m_\chi \right)} \,,
\label{eq:sigma_hier_uR}
\end{align}
where we defined $\xi_{\rm SD/SI} \left(m_\chi \right) \equiv \sigma_{\rm SD/SI}^{{\rm max}, gg} \left(m_\chi \right) / \sigma_{\rm SD/SI}^{{\rm max}, ZZ} \left(m_\chi \right)$. In \EquTworef{eq:sigma_deg_uR}{eq:sigma_hier_uR}, the second term in each denominator parametrizes the contribution of the annihilation channels containing a $Z$ boson in the final state. From the results in Appendix \ref{ap:ZZ-gg} it follows that $\xi_{\rm SD/SI} \left(m_\chi \right)$ is a decreasing function of $m_\chi$, hence the relative contribution to the limits of annihilation channels involving a $Z$ boson gets smaller for larger dark matter masses $m_\chi$. Concretely, for $m_\chi = 10$ TeV, the limits dominantly arise from the channels $u_R \bar{u}_R g$ ($gg$) in the degenerate (hierarchical) case, and consequently the ratio of $\sigma^{\rm max, deg}_{\rm SD/SI}$ and $\sigma^{\rm max, hier}_{\rm SD/SI}$ is $\simeq 2$, as it is in the case of couplings to $e_R$. 
For smaller $m_\chi$, also the annihilation channels $u_R \bar{u}_R Z$ and $\gamma Z, ZZ$ contribute significantly to the limits, and it follows from \EquTworef{eq:sigma_deg_uR}{eq:sigma_hier_uR} that due to ${\rm BF}_{u_R \bar{u}_R Z} > {\rm BF}_{ZZ}+{\rm BF}_{\gamma Z}/2$, the upper limits in the degenerate case improve to a larger extent than in the hierarchical case when lowering $m_\chi$. Hence, as shown in \Figref{fig:limits-toy-model}, the upper limits are even more insensitive to the mass splitting $m_\eta/m_\chi$ at $m_\chi \simeq $ 1 TeV than at $\simeq 10 $ TeV. Similar arguments also apply in the case of couplings to $d_R$.

In \Figref{fig:limits-toy-model}, we also show for comparison the limits on the spin-dependent interaction cross sections from COUPP and SIMPLE, and on the spin-independent interaction cross sections from XENON100 and LUX. For couplings to leptons, the limits on the spin-dependent cross section are significantly stronger than the direct detection limits, while for quarks they are weaker, since in this case the dark matter particle annihilates mostly into final states involving gluons which have a much larger branching fraction than those involving weak gauge bosons. On the other hand, the limits on the spin-independent cross section from IceCube are much weaker, at least one order of magnitude than the direct detection limits for couplings to $e_R$ and at least two orders of magnitude for couplings to $u_R$.

\section{Conclusions}
\label{sec:Conclusions}

We have investigated the impact of higher order annihilation processes in the generation of a high energy neutrino flux from dark matter annihilations in the Sun in scenarios where the dark matter particle couples to the electron, the muon or to the light quark. We have argued that, while the annihilation into a fermion-antifermion pair generates only MeV neutrinos, the associated emission of a gauge boson off the final state and the loop induced annihilation into two gauge bosons do generate a high energy neutrino flux, thus opening the possibility of probing these scenarios at IceCube. 

We have first adopted a model independent approach where the annihilation into the fermion-antifermion is induced by a contact interaction and we have considered two limiting scenarios depending on which is the most important source of high energy neutrinos, whether the loop annihilations or the final state radiation of gauge bosons. In both cases, we have derived limits on the spin-dependent and spin-independent dark matter interaction cross sections with protons under the assumption that captures and annihilations are in equilibrium in the solar interior. We have found fairly stringent limits for the spin-dependent scattering cross section for dark matter masses between 100 GeV and 10 TeV which are complementary to the limits from the direct search experiments COUPP and SIMPLE and, in the case of coupling to leptons, stronger. 

Lastly, we have carefully analyzed the neutrino signals in a toy model consisting in a Majorana fermion as dark matter particle that couples to a right-handed electron or first generation quark  via a Yukawa coupling with a scalar. The branching ratios of the different processes are calculable in this model: the two-to-two annihilation into a fermion-antifermion pair is helicity- and velocity-suppressed, and always negligible, while the two-to-three and loop induced annihilations into weak gauge bosons have a sizable branching fraction, their relative importance being dependent on the parameters of the model. 
More specifically, the two-to-three process dominates when the scalar that mediates the interaction is degenerate in mass with the dark matter particle, while the loop process dominates when there is a large mass hierarchy between them. We have found that the limits on the interaction cross section depend mildly, at most by a factor of two, on the mass of the scalar particle that mediates the interaction. Furthermore, we have found stringent limits on the spin-dependent interaction cross section for the model where the dark matter particle couples to right-handed electrons, which are stronger than the limits set by direct search experiments.

\vspace{0.5cm}
\section*{Acknowledgements}
This work was partially supported by the DFG cluster of excellence ``Origin and Structure of the Universe,'' the TUM Graduate School and the Studienstiftung des Deutschen Volkes.

\FloatBarrier 

\appendix

\section{Fit coefficients for the neutrino spectra from final state radiation}
\label{ap:parametrization}

In this appendix, we provide fit functions for the (anti-)neutrino spectra originating from annihilations of dark matter particles with mass $m_{\rm DM}$ that couple to a fermion-antifermion pair through a contact interaction. Concretely, the radiation of weak gauge bosons and/or gluons off the fermion of the final state produces a (anti-)neutrino spectrum at Earth with a differential spectrum per annihilation that we fit by the function 
\begin{align}
\log_{10} \left( \frac{\mathrm{d} \Phi}{\mathrm{d} E} \right) = \Theta \left( E_0 - E \right) \sum_{i=0}^{4} a_i \left( \log_{10}  E   \right)^i + \Theta \left( E - E_0 \right) \sum_{i=0}^{4} b_i \left( \log_{10}  E - \log_{10} E_0   \right)^i \,.
\end{align}
The parameters of the fit $a_i$, $b_i$ and $E_0$ are given in the tables shown below, for couplings to $e_{R/L},\mu_{R/L},u_{R/L},d_{R/L},c_{R/L}$ and $s_{R/L}$. Here $E$ is the (anti-)neutrino energy in GeV and $\frac{\mathrm{d} \Phi}{\mathrm{d} E}$ in units of $\mathrm{cm}^{-2} \, \mathrm{GeV}^{-1}$. All lines not containing an entry for $E_0$ correspond to a fit with $E_0 >m_{\rm DM}>E$, and hence only the parameters $a_i$ are shown in these cases. 

Furthermore, we show in \Figref{fig:Fits} a sample of neutrino spectra at Earth, comparing the full numerical results with the corresponding fit functions. As apparent from the plots, the fit functions provide a good description of the neutrino spectra for $10\GeV < E< 2\TeV$, which is the energy range relevant for neutrino detection (for $E>2\TeV$ neutrino energy losses due to interactions with solar matter become very efficient, resulting in very suppressed fluxes~\cite{Ritz:1987mh}).

\FloatBarrier

\begin{figure}[h!]
\begin{center}
\includegraphics[width=0.49\textwidth]{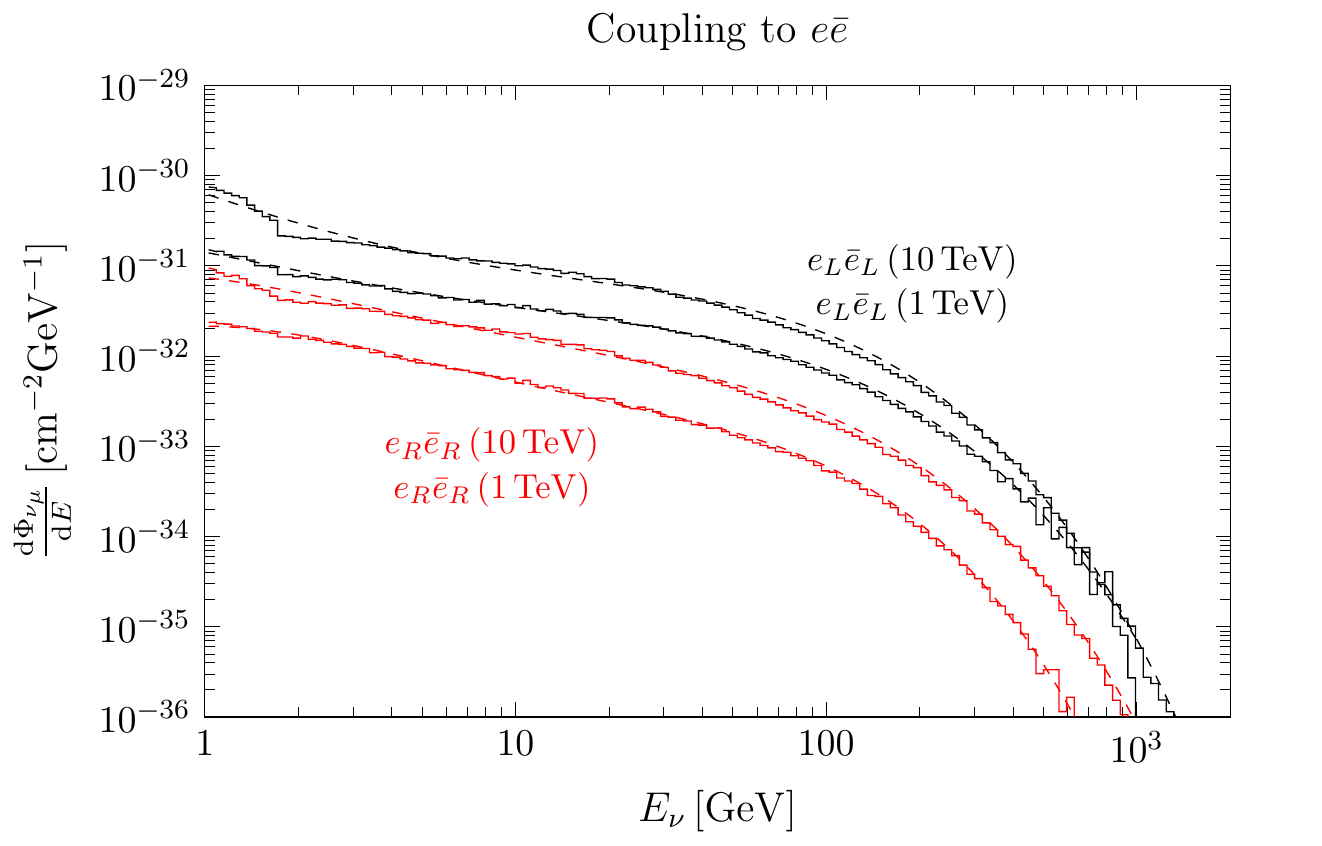} \\
\includegraphics[width=0.49\textwidth]{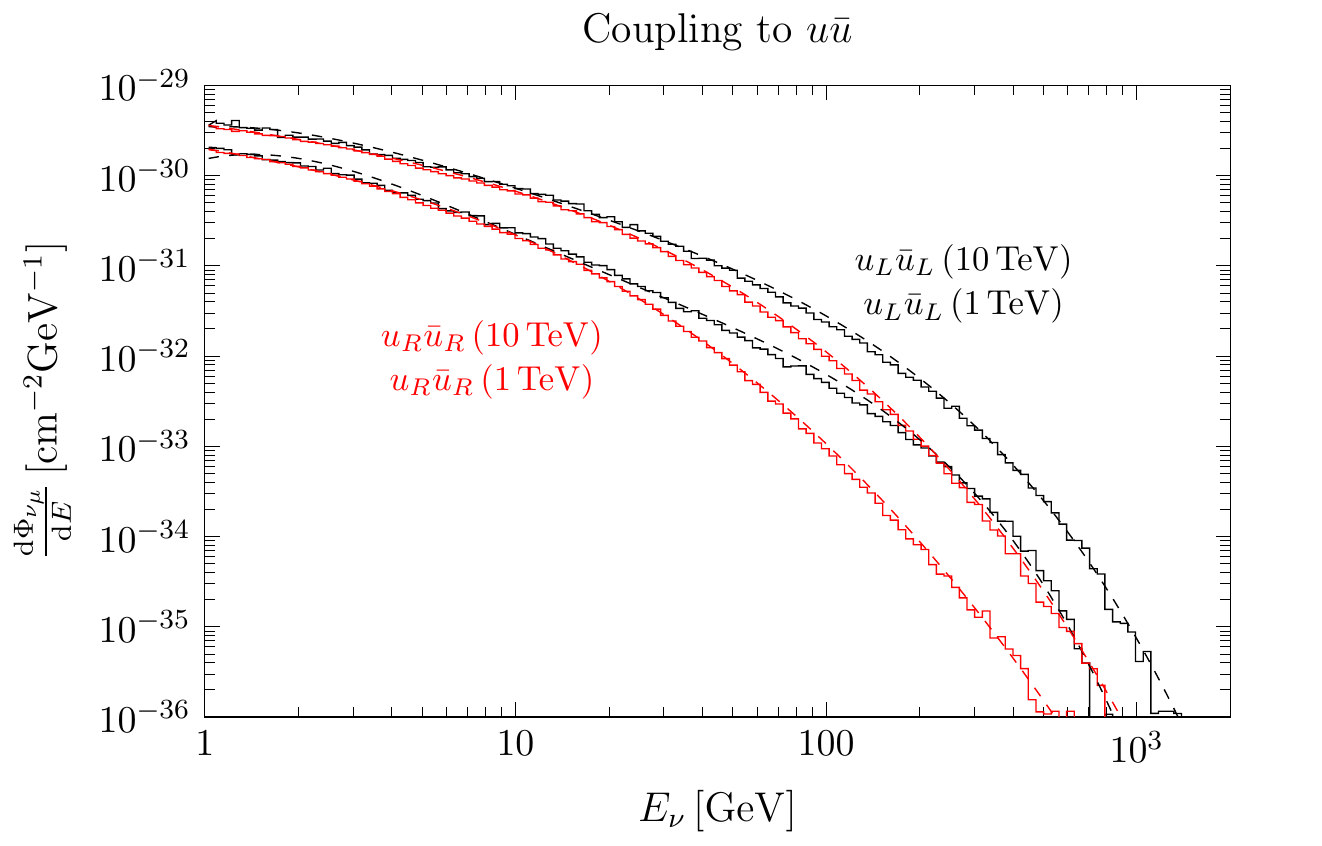}
\includegraphics[width=0.49\textwidth]{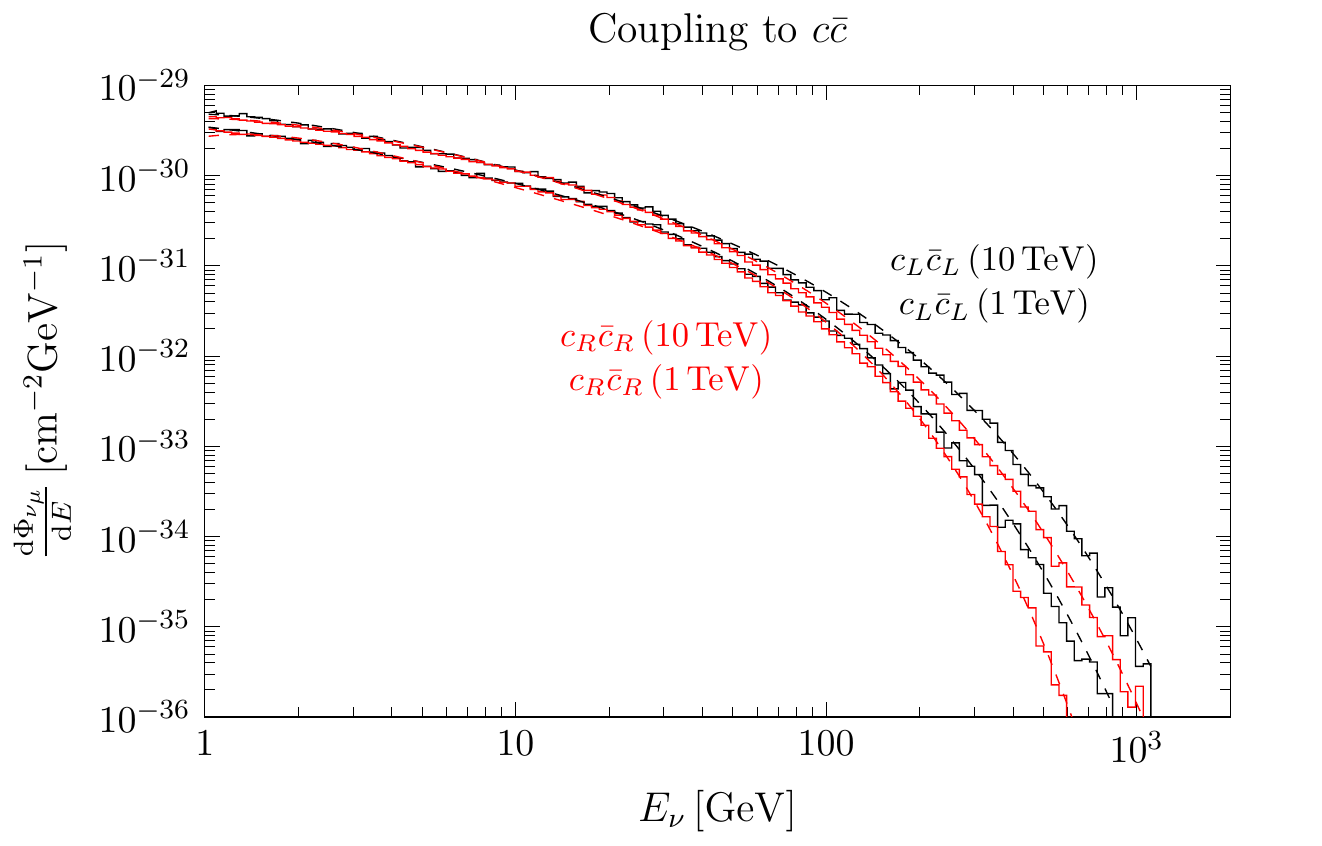}
\caption{Comparison of neutrino spectra at Earth from final state radiation calculated using the full numerical analysis (histograms) and the fitting functions presented in Appendix \ref{ap:parametrization} (dashed lines), for different dark matter couplings and for $m_{\rm DM}=1\TeV$ and 10 TeV.}
\label{fig:Fits}
\end{center}
\end{figure}

{\renewcommand{\arraystretch}{0.3} 
 \begin{table}[htbp] 
 \footnotesize 
 \begin{tabular}{cccccccc} 
\toprule 
\multicolumn{8}{c}{Coupling to $e_R\bar{e}_R$ or $\mu_R\bar{\mu}_R$} \\ [0.2cm] 
$m_{\mathrm{DM}}$ & $E_0$ & & $a_0 / b_0$ & $a_1 / b_1$ & $a_2 / b_2$ & $a_3 / b_3$ & $a_4 / b_4$\\[0.1cm] 
\toprule 
\multirow{2}{*}{100} & \multirow{2}{*}{35}& $\nu_\mu$ &-32.57/-34.6&-0.717/-1.727&0.032/19.42&-0.635/-80.15&0.216/-35.8 \\ \cmidrule{4-8}&& $\bar{\nu}_\mu$ &-32.5/-34.54&-0.981/-0.853&1.072/2.74&-1.809/7.754&0.609/-157.7 \\ \midrule 
\multirow{2}{*}{200} & \multirow{2}{*}{100}& $\nu_\mu$ &-32.29/-34.09&0.193/-3.431&-2.371/-42.81&1.854/459.9&-0.475/-1329 \\ \cmidrule{4-8}&& $\bar{\nu}_\mu$ &-32.27/-34.18&0.234/7.727&-2.095/-206.6&1.425/1342&-0.326/-2871 \\ \midrule 
\multirow{2}{*}{300} & \multirow{2}{*}{170}& $\nu_\mu$ &-32.11/-34.48&0.134/-5.456&-1.964/63.56&1.522/-497.&-0.394/701.7 \\ \cmidrule{4-8}&& $\bar{\nu}_\mu$ &-32.09/-34.27&0.278/0.344&-2.144/-97.07&1.544/785.4&-0.373/-2233 \\ \midrule 
\multirow{2}{*}{500} & \multirow{2}{*}{}& $\nu_\mu$ &-31.91&0.304&-2.127&1.559&-0.384 \\ \cmidrule{4-8}&& $\bar{\nu}_\mu$ &-31.93&0.604&-2.571&1.768&-0.408 \\ \midrule 
\multirow{2}{*}{800} & \multirow{2}{*}{}& $\nu_\mu$ &-31.74&-0.015&-1.32&0.964&-0.248 \\ \cmidrule{4-8}&& $\bar{\nu}_\mu$ &-31.76&0.332&-1.886&1.253&-0.287 \\ \midrule 
\multirow{2}{*}{1000} & \multirow{2}{*}{}& $\nu_\mu$ &-31.67&0.001&-1.338&0.961&-0.243 \\ \cmidrule{4-8}&& $\bar{\nu}_\mu$ &-31.7&0.394&-1.933&1.248&-0.279 \\ \midrule 
\multirow{2}{*}{3000} & \multirow{2}{*}{}& $\nu_\mu$ &-31.44&0.251&-1.722&1.161&-0.272 \\ \cmidrule{4-8}&& $\bar{\nu}_\mu$ &-31.4&0.043&-1.264&0.812&-0.185 \\ \midrule 
\multirow{2}{*}{5000} & \multirow{2}{*}{}& $\nu_\mu$ &-31.25&-0.366&-0.675&0.553&-0.159 \\ \cmidrule{4-8}&& $\bar{\nu}_\mu$ &-31.29&0.029&-1.238&0.792&-0.18 \\ \midrule 
\multirow{2}{*}{10000} & \multirow{2}{*}{}& $\nu_\mu$ &-31.13&-0.413&-0.628&0.534&-0.154 \\ \cmidrule{4-8}&& $\bar{\nu}_\mu$ &-31.17&0.011&-1.25&0.807&-0.182 \\ \midrule 
\bottomrule 
\end{tabular} 
 \end{table} } 
 
{\renewcommand{\arraystretch}{0.3} 
 \begin{table}[htbp] 
 \footnotesize 
 \begin{tabular}{cccccccc} 
\toprule 
\multicolumn{8}{c}{Coupling to $e_L\bar{e}_L$ or $\mu_L \bar{\mu}_L$} \\ [0.2cm] 
$m_{\mathrm{DM}}$ & $E_0$ & & $a_0 / b_0$ & $a_1 / b_1$ & $a_2 / b_2$ & $a_3 / b_3$ & $a_4 / b_4$\\[0.1cm] 
\toprule 
\multirow{2}{*}{100} & \multirow{2}{*}{35}& $\nu_\mu$ &-32.41/-32.93&-0.668/-1.574&0.824/10.52&-2.042/12.54&1.071/-169.2 \\ \cmidrule{4-8}&& $\bar{\nu}_\mu$ &-32.34/-32.95&-1.075/-1.687&1.98/24.42&-3.21/-74.35&1.438/-29.93 \\ \midrule 
\multirow{2}{*}{200} & \multirow{2}{*}{100}& $\nu_\mu$ &-31.68/-32.7&-0.262/8.437&-1.264/-182.3&1.327/1223&-0.374/-2708 \\ \cmidrule{4-8}&& $\bar{\nu}_\mu$ &-31.68/-32.57&-0.228/7.901&-1.175/-161.9&1.125/1084&-0.294/-2414 \\ \midrule 
\multirow{2}{*}{300} & \multirow{2}{*}{170}& $\nu_\mu$ &-31.36/-32.81&-0.731/16.12&-0.053/-440.9&0.321/3539&-0.124/-8949 \\ \cmidrule{4-8}&& $\bar{\nu}_\mu$ &-31.38/-32.6&-0.64/13.54&-0.036/-372.9&0.165/3019&-0.055/-7723 \\ \midrule 
\multirow{2}{*}{500} & \multirow{2}{*}{}& $\nu_\mu$ &-31.15&-0.3&-0.808&0.754&-0.207 \\ \cmidrule{4-8}&& $\bar{\nu}_\mu$ &-31.18&-0.063&-1.139&0.879&-0.212 \\ \midrule 
\multirow{2}{*}{800} & \multirow{2}{*}{}& $\nu_\mu$ &-30.93&-0.64&-0.156&0.309&-0.114 \\ \cmidrule{4-8}&& $\bar{\nu}_\mu$ &-30.93&-0.578&-0.218&0.281&-0.09 \\ \midrule 
\multirow{2}{*}{1000} & \multirow{2}{*}{}& $\nu_\mu$ &-30.85&-0.667&-0.111&0.276&-0.108 \\ \cmidrule{4-8}&& $\bar{\nu}_\mu$ &-30.84&-0.689&-0.043&0.177&-0.071 \\ \midrule 
\multirow{2}{*}{3000} & \multirow{2}{*}{}& $\nu_\mu$ &-30.51&-0.814&-0.111&0.336&-0.128 \\ \cmidrule{4-8}&& $\bar{\nu}_\mu$ &-30.5&-0.798&-0.091&0.259&-0.095 \\ \midrule 
\multirow{2}{*}{5000} & \multirow{2}{*}{}& $\nu_\mu$ &-30.38&-0.836&-0.182&0.402&-0.143 \\ \cmidrule{4-8}&& $\bar{\nu}_\mu$ &-30.33&-1.075&0.193&0.154&-0.083 \\ \midrule 
\multirow{2}{*}{10000} & \multirow{2}{*}{}& $\nu_\mu$ &-30.2&-1.175&0.227&0.214&-0.114 \\ \cmidrule{4-8}&& $\bar{\nu}_\mu$ &-30.18&-1.118&0.134&0.213&-0.096 \\ \midrule 
\bottomrule 
\end{tabular} 
 \end{table} } 
 
{\renewcommand{\arraystretch}{0.3} 
 \begin{table}[htbp] 
 \footnotesize 
 \begin{tabular}{cccccccc} 
\toprule 
\multicolumn{8}{c}{Coupling to $u_R\bar{u}_R$} \\ [0.2cm] 
$m_{\mathrm{DM}}$ & $E_0$ & & $a_0 / b_0$ & $a_1 / b_1$ & $a_2 / b_2$ & $a_3 / b_3$ & $a_4 / b_4$\\[0.1cm] 
\toprule 
\multirow{2}{*}{100} & \multirow{2}{*}{35}& $\nu_\mu$ &-30.13/-33.79&-0.33/-9.258&-2.272/27.06&1.875/-87.95&-0.834/68.4 \\ \cmidrule{4-8}&& $\bar{\nu}_\mu$ &-30.1/-33.41&-0.564/-28.64&-1.335/262.3&0.851/-1056&-0.499/1329 \\ \midrule 
\multirow{2}{*}{200} & \multirow{2}{*}{100}& $\nu_\mu$ &-29.91/-34.54&-1.097/-5.457&0.495/53.31&-0.907/-323.9&0.171/409.7 \\ \cmidrule{4-8}&& $\bar{\nu}_\mu$ &-29.89/-34.44&-1.15/0.622&0.869/-81.97&-1.341/603.9&0.308/-1490 \\ \midrule 
\multirow{2}{*}{300} & \multirow{2}{*}{170}& $\nu_\mu$ &-29.82/-34.83&-1.293/-2.185&1.101/-46.24&-1.322/462.4&0.284/-1642 \\ \cmidrule{4-8}&& $\bar{\nu}_\mu$ &-29.81/-34.71&-1.189/0.873&1.095/-132.4&-1.437/1264&0.336/-3835 \\ \midrule 
\multirow{2}{*}{500} & \multirow{2}{*}{}& $\nu_\mu$ &-29.8&-0.593&-0.292&-0.212&0.023 \\ \cmidrule{4-8}&& $\bar{\nu}_\mu$ &-29.83&-0.225&-0.862&0.084&-0.02 \\ \midrule 
\multirow{2}{*}{800} & \multirow{2}{*}{}& $\nu_\mu$ &-29.75&-0.398&-0.596&0.055&-0.038 \\ \cmidrule{4-8}&& $\bar{\nu}_\mu$ &-29.8&0.112&-1.381&0.453&-0.095 \\ \midrule 
\multirow{2}{*}{1000} & \multirow{2}{*}{}& $\nu_\mu$ &-29.7&-0.614&-0.153&-0.205&0.013 \\ \cmidrule{4-8}&& $\bar{\nu}_\mu$ &-29.75&0.002&-1.156&0.341&-0.075 \\ \midrule 
\multirow{2}{*}{3000} & \multirow{2}{*}{}& $\nu_\mu$ &-29.51&-0.952&0.501&-0.47&0.047 \\ \cmidrule{4-8}&& $\bar{\nu}_\mu$ &-29.61&-0.107&-0.828&0.226&-0.06 \\ \midrule 
\multirow{2}{*}{5000} & \multirow{2}{*}{}& $\nu_\mu$ &-29.48&-0.742&0.19&-0.265&0.008 \\ \cmidrule{4-8}&& $\bar{\nu}_\mu$ &-29.54&-0.169&-0.669&0.153&-0.048 \\ \midrule 
\multirow{2}{*}{10000} & \multirow{2}{*}{}& $\nu_\mu$ &-29.44&-0.461&-0.236&0.007&-0.044 \\ \cmidrule{4-8}&& $\bar{\nu}_\mu$ &-29.44&-0.329&-0.379&0.034&-0.033 \\ \midrule 
\bottomrule 
\end{tabular} 
 \end{table} } 
 
{\renewcommand{\arraystretch}{0.3} 
 \begin{table}[htbp] 
 \footnotesize 
 \begin{tabular}{cccccccc} 
\toprule 
\multicolumn{8}{c}{Coupling to $u_L\bar{u}_L$} \\ [0.2cm] 
$m_{\mathrm{DM}}$ & $E_0$ & & $a_0 / b_0$ & $a_1 / b_1$ & $a_2 / b_2$ & $a_3 / b_3$ & $a_4 / b_4$\\[0.1cm] 
\toprule 
\multirow{2}{*}{100} & \multirow{2}{*}{35}& $\nu_\mu$ &-30.07/-33.5&-1.238/5.293&0.935/-77.97&-2.003/346.1&0.661/-566.8 \\ \cmidrule{4-8}&& $\bar{\nu}_\mu$ &-30.06/-33.18&-1.125/-4.926&0.706/27.77&-1.728/-34.71&0.538/-132.2 \\ \midrule 
\multirow{2}{*}{200} & \multirow{2}{*}{100}& $\nu_\mu$ &-29.95/-33.37&-0.586/10.48&-0.589/-233.3&-0.28/1433&0.161/-2969 \\ \cmidrule{4-8}&& $\bar{\nu}_\mu$ &-29.93/-33.17&-0.626/4.874&-0.331/-140.6&-0.545/914.2&0.24/-1999 \\ \midrule 
\multirow{2}{*}{300} & \multirow{2}{*}{170}& $\nu_\mu$ &-29.9/-33.52&-0.284/3.336&-1.265/-182.6&0.466/1477&-0.07/-3880 \\ \cmidrule{4-8}&& $\bar{\nu}_\mu$ &-29.87/-33.31&-0.354/-0.912&-0.933/-69.18&0.12/654.6&0.033/-2105 \\ \midrule 
\multirow{2}{*}{500} & \multirow{2}{*}{}& $\nu_\mu$ &-29.9&0.658&-3.051&1.788&-0.378 \\ \cmidrule{4-8}&& $\bar{\nu}_\mu$ &-29.92&0.867&-3.271&1.828&-0.363 \\ \midrule 
\multirow{2}{*}{800} & \multirow{2}{*}{}& $\nu_\mu$ &-29.85&0.677&-2.828&1.602&-0.332 \\ \cmidrule{4-8}&& $\bar{\nu}_\mu$ &-29.86&0.986&-3.261&1.783&-0.346 \\ \midrule 
\multirow{2}{*}{1000} & \multirow{2}{*}{}& $\nu_\mu$ &-29.82&0.754&-2.845&1.58&-0.322 \\ \cmidrule{4-8}&& $\bar{\nu}_\mu$ &-29.82&0.918&-3.028&1.613&-0.308 \\ \midrule 
\multirow{2}{*}{3000} & \multirow{2}{*}{}& $\nu_\mu$ &-29.59&0.055&-1.394&0.718&-0.16 \\ \cmidrule{4-8}&& $\bar{\nu}_\mu$ &-29.65&0.6&-2.19&1.094&-0.207 \\ \midrule 
\multirow{2}{*}{5000} & \multirow{2}{*}{}& $\nu_\mu$ &-29.53&0.012&-1.244&0.631&-0.144 \\ \cmidrule{4-8}&& $\bar{\nu}_\mu$ &-29.55&0.341&-1.707&0.829&-0.161 \\ \midrule 
\multirow{2}{*}{10000} & \multirow{2}{*}{}& $\nu_\mu$ &-29.46&0.11&-1.303&0.662&-0.15 \\ \cmidrule{4-8}&& $\bar{\nu}_\mu$ &-29.43&0.049&-1.153&0.529&-0.11 \\ \midrule 
\bottomrule 
\end{tabular} 
 \end{table} } 
 
 {\renewcommand{\arraystretch}{0.3} 
 \begin{table}[htbp] 
 \footnotesize 
 \begin{tabular}{cccccccc} 
\toprule 
\multicolumn{8}{c}{Coupling to $d_R\bar{d}_R$} \\ [0.2cm] 
$m_{\mathrm{DM}}$ & $E_0$ & & $a_0 / b_0$ & $a_1 / b_1$ & $a_2 / b_2$ & $a_3 / b_3$ & $a_4 / b_4$\\[0.1cm] 
\toprule 
\multirow{2}{*}{100} & \multirow{2}{*}{35}& $\nu_\mu$ &-30.13/-34.28&-0.302/11.15&-2.39/-234.8&2.008/1050&-0.878/-1553 \\ \cmidrule{4-8}&& $\bar{\nu}_\mu$ &-30.09/-33.74&-0.7/-13.63&-0.885/114.7&0.361/-706.4&-0.333/1320 \\ \midrule 
\multirow{2}{*}{200} & \multirow{2}{*}{100}& $\nu_\mu$ &-29.95/-34.94&-0.621/-9.897&-0.871/89.&0.447/-515.6&-0.261/919.1 \\ \cmidrule{4-8}&& $\bar{\nu}_\mu$ &-29.92/-34.94&-0.732/-7.376&-0.364/51.11&-0.101/-174.9&-0.09/-36.74 \\ \midrule 
\multirow{2}{*}{300} & \multirow{2}{*}{170}& $\nu_\mu$ &-29.83/-35.47&-1.062/0.554&0.403/-162.6&-0.608/1566&0.053/-4652 \\ \cmidrule{4-8}&& $\bar{\nu}_\mu$ &-29.83/-35.3&-0.967/1.838&0.401/-117.2&-0.727/850.9&0.108/-2229 \\ \midrule 
\multirow{2}{*}{500} & \multirow{2}{*}{}& $\nu_\mu$ &-29.76&-0.945&0.34&-0.552&0.063 \\ \cmidrule{4-8}&& $\bar{\nu}_\mu$ &-29.76&-0.761&0.114&-0.479&0.067 \\ \midrule 
\multirow{2}{*}{800} & \multirow{2}{*}{}& $\nu_\mu$ &-29.71&-0.779&0.074&-0.31&0.012 \\ \cmidrule{4-8}&& $\bar{\nu}_\mu$ &-29.73&-0.499&-0.299&-0.162&0.003 \\ \midrule 
\multirow{2}{*}{1000} & \multirow{2}{*}{}& $\nu_\mu$ &-29.65&-0.988&0.457&-0.513&0.05 \\ \cmidrule{4-8}&& $\bar{\nu}_\mu$ &-29.69&-0.597&-0.081&-0.28&0.027 \\ \midrule 
\multirow{2}{*}{3000} & \multirow{2}{*}{}& $\nu_\mu$ &-29.56&-0.58&-0.15&-0.066&-0.038 \\ \cmidrule{4-8}&& $\bar{\nu}_\mu$ &-29.52&-0.717&0.127&-0.26&0.01 \\ \midrule 
\multirow{2}{*}{5000} & \multirow{2}{*}{}& $\nu_\mu$ &-29.53&-0.415&-0.419&0.136&-0.08 \\ \cmidrule{4-8}&& $\bar{\nu}_\mu$ &-29.43&-0.982&0.588&-0.485&0.048 \\ \midrule 
\multirow{2}{*}{10000} & \multirow{2}{*}{}& $\nu_\mu$ &-29.44&-0.444&-0.32&0.094&-0.071 \\ \cmidrule{4-8}&& $\bar{\nu}_\mu$ &-29.44&-0.34&-0.408&0.086&-0.052 \\ \midrule 
\bottomrule 
\end{tabular} 
 \end{table} } 
 
{\renewcommand{\arraystretch}{0.3} 
 \begin{table}[htbp] 
 \footnotesize 
 \begin{tabular}{cccccccc} 
\toprule 
\multicolumn{8}{c}{Coupling to $d_L\bar{d}_L$} \\ [0.2cm] 
$m_{\mathrm{DM}}$ & $E_0$ & & $a_0 / b_0$ & $a_1 / b_1$ & $a_2 / b_2$ & $a_3 / b_3$ & $a_4 / b_4$\\[0.1cm] 
\toprule 
\multirow{2}{*}{100} & \multirow{2}{*}{35}& $\nu_\mu$ &-30.11/-33.52&-0.781/8.822&-0.401/-129.&-0.623/607.6&0.201/-997.9 \\ \cmidrule{4-8}&& $\bar{\nu}_\mu$ &-30.07/-33.56&-1.048/8.903&0.477/-116.8&-1.476/523.7&0.457/-842.6 \\ \midrule 
\multirow{2}{*}{200} & \multirow{2}{*}{100}& $\nu_\mu$ &-29.96/-33.43&-0.412/13.29&-1.148/-283.9&0.271/1681&0.005/-3330 \\ \cmidrule{4-8}&& $\bar{\nu}_\mu$ &-29.96/-33.11&-0.415/-0.143&-0.79/-71.39&-0.202/516.9&0.162/-1276 \\ \midrule 
\multirow{2}{*}{300} & \multirow{2}{*}{170}& $\nu_\mu$ &-29.9/-33.56&-0.086/-2.218&-1.819/-51.79&0.956/379.&-0.198/-1170 \\ \cmidrule{4-8}&& $\bar{\nu}_\mu$ &-29.89/-33.37&-0.082/0.571&-1.614/-128.1&0.693/1110&-0.113/-3213 \\ \midrule 
\multirow{2}{*}{500} & \multirow{2}{*}{}& $\nu_\mu$ &-29.94&0.925&-3.645&2.263&-0.494 \\ \cmidrule{4-8}&& $\bar{\nu}_\mu$ &-29.92&0.924&-3.491&2.063&-0.429 \\ \midrule 
\multirow{2}{*}{800} & \multirow{2}{*}{}& $\nu_\mu$ &-29.84&0.767&-3.071&1.817&-0.386 \\ \cmidrule{4-8}&& $\bar{\nu}_\mu$ &-29.88&1.233&-3.788&2.174&-0.434 \\ \midrule 
\multirow{2}{*}{1000} & \multirow{2}{*}{}& $\nu_\mu$ &-29.79&0.545&-2.551&1.469&-0.313 \\ \cmidrule{4-8}&& $\bar{\nu}_\mu$ &-29.81&0.906&-3.095&1.721&-0.34 \\ \midrule 
\multirow{2}{*}{3000} & \multirow{2}{*}{}& $\nu_\mu$ &-29.61&0.33&-1.939&1.072&-0.229 \\ \cmidrule{4-8}&& $\bar{\nu}_\mu$ &-29.64&0.609&-2.224&1.133&-0.217 \\ \midrule 
\multirow{2}{*}{5000} & \multirow{2}{*}{}& $\nu_\mu$ &-29.52&0.073&-1.408&0.761&-0.172 \\ \cmidrule{4-8}&& $\bar{\nu}_\mu$ &-29.53&0.254&-1.573&0.771&-0.153 \\ \midrule 
\multirow{2}{*}{10000} & \multirow{2}{*}{}& $\nu_\mu$ &-29.46&0.109&-1.347&0.718&-0.164 \\ \cmidrule{4-8}&& $\bar{\nu}_\mu$ &-29.5&0.584&-2.02&1.007&-0.193 \\ \midrule 
\bottomrule 
\end{tabular} 
 \end{table} } 
 
{\renewcommand{\arraystretch}{0.3} 
 \begin{table}[htbp] 
 \footnotesize 
 \begin{tabular}{cccccccc} 
\toprule 
\multicolumn{8}{c}{Coupling to $c_R\bar{c}_R$} \\ [0.2cm] 
$m_{\mathrm{DM}}$ & $E_0$ & & $a_0 / b_0$ & $a_1 / b_1$ & $a_2 / b_2$ & $a_3 / b_3$ & $a_4 / b_4$\\[0.1cm] 
\toprule 
\multirow{2}{*}{100} & \multirow{2}{*}{35}& $\nu_\mu$ &-29.47/-31.16&0.049/-7.758&-1.507/39.45&1.522/-193.2&-0.679/198.1 \\ \cmidrule{4-8}&& $\bar{\nu}_\mu$ &-29.44/-31.24&-0.18/-3.743&-0.576/-2.718&0.5/-25.58&-0.345/-23.26 \\ \midrule 
\multirow{2}{*}{200} & \multirow{2}{*}{100}& $\nu_\mu$ &-29.53/-32.61&0.033/-17.38&-1.378/171.2&1.255/-1243&-0.483/2659 \\ \cmidrule{4-8}&& $\bar{\nu}_\mu$ &-29.51/-32.58&-0.018/-6.62&-1.026/-28.92&0.845/16.39&-0.351/99.76 \\ \midrule 
\multirow{2}{*}{300} & \multirow{2}{*}{170}& $\nu_\mu$ &-29.54/-33.67&0.034/-13.1&-1.333/191.7&1.12/-2276&-0.396/6533 \\ \cmidrule{4-8}&& $\bar{\nu}_\mu$ &-29.54/-33.51&0.114/0.853&-1.309/-321.2&0.991/2585&-0.342/-6642 \\ \midrule 
\multirow{2}{*}{500} & \multirow{2}{*}{}& $\nu_\mu$ &-29.58&0.444&-2.14&1.588&-0.455 \\ \cmidrule{4-8}&& $\bar{\nu}_\mu$ &-29.63&0.991&-3.072&2.127&-0.549 \\ \midrule 
\multirow{2}{*}{800} & \multirow{2}{*}{}& $\nu_\mu$ &-29.57&0.352&-1.82&1.24&-0.34 \\ \cmidrule{4-8}&& $\bar{\nu}_\mu$ &-29.58&0.598&-2.171&1.395&-0.354 \\ \midrule 
\multirow{2}{*}{1000} & \multirow{2}{*}{}& $\nu_\mu$ &-29.57&0.41&-1.871&1.218&-0.321 \\ \cmidrule{4-8}&& $\bar{\nu}_\mu$ &-29.53&0.201&-1.393&0.861&-0.234 \\ \midrule 
\multirow{2}{*}{3000} & \multirow{2}{*}{}& $\nu_\mu$ &-29.46&-0.019&-0.946&0.532&-0.156 \\ \cmidrule{4-8}&& $\bar{\nu}_\mu$ &-29.45&0.063&-1.018&0.523&-0.138 \\ \midrule 
\multirow{2}{*}{5000} & \multirow{2}{*}{}& $\nu_\mu$ &-29.38&-0.322&-0.443&0.247&-0.104 \\ \cmidrule{4-8}&& $\bar{\nu}_\mu$ &-29.41&-0.004&-0.862&0.405&-0.11 \\ \midrule 
\multirow{2}{*}{10000} & \multirow{2}{*}{}& $\nu_\mu$ &-29.37&-0.013&-0.881&0.454&-0.133 \\ \cmidrule{4-8}&& $\bar{\nu}_\mu$ &-29.39&0.178&-1.099&0.508&-0.123 \\ \midrule 
\bottomrule 
\end{tabular} 
 \end{table} } 
 
{\renewcommand{\arraystretch}{0.3} 
 \begin{table}[htbp] 
 \footnotesize 
 \begin{tabular}{cccccccc} 
\toprule 
\multicolumn{8}{c}{Coupling to $c_L\bar{c}_L$} \\ [0.2cm] 
$m_{\mathrm{DM}}$ & $E_0$ & & $a_0 / b_0$ & $a_1 / b_1$ & $a_2 / b_2$ & $a_3 / b_3$ & $a_4 / b_4$\\[0.1cm] 
\toprule 
\multirow{2}{*}{100} & \multirow{2}{*}{35}& $\nu_\mu$ &-29.48/-31.23&0.12/-4.346&-1.674/1.888&1.673/-43.55&-0.723/14.61 \\ \cmidrule{4-8}&& $\bar{\nu}_\mu$ &-29.46/-31.2&-0.202/-5.676&-0.416/23.95&0.318/-149.4&-0.286/172. \\ \midrule 
\multirow{2}{*}{200} & \multirow{2}{*}{100}& $\nu_\mu$ &-29.5/-32.43&-0.188/-15.02&-0.895/130.8&0.872/-741.&-0.381/1298 \\ \cmidrule{4-8}&& $\bar{\nu}_\mu$ &-29.52/-32.45&0.046/-6.664&-1.184/-4.09&0.974/37.03&-0.383/-74.41 \\ \midrule 
\multirow{2}{*}{300} & \multirow{2}{*}{170}& $\nu_\mu$ &-29.51/-33.32&-0.268/-2.864&-0.607/-36.73&0.524/234.5&-0.24/-928.2 \\ \cmidrule{4-8}&& $\bar{\nu}_\mu$ &-29.52/-33.25&-0.105/13.39&-0.753/-395.2&0.505/2855&-0.206/-6773 \\ \midrule 
\multirow{2}{*}{500} & \multirow{2}{*}{}& $\nu_\mu$ &-29.51&-0.272&-0.537&0.368&-0.159 \\ \cmidrule{4-8}&& $\bar{\nu}_\mu$ &-29.53&0.025&-1.003&0.606&-0.191 \\ \midrule 
\multirow{2}{*}{800} & \multirow{2}{*}{}& $\nu_\mu$ &-29.47&-0.44&-0.252&0.166&-0.102 \\ \cmidrule{4-8}&& $\bar{\nu}_\mu$ &-29.54&0.221&-1.305&0.724&-0.188 \\ \midrule 
\multirow{2}{*}{1000} & \multirow{2}{*}{}& $\nu_\mu$ &-29.46&-0.387&-0.352&0.216&-0.105 \\ \cmidrule{4-8}&& $\bar{\nu}_\mu$ &-29.55&0.335&-1.443&0.771&-0.187 \\ \midrule 
\multirow{2}{*}{3000} & \multirow{2}{*}{}& $\nu_\mu$ &-29.42&-0.245&-0.565&0.292&-0.1 \\ \cmidrule{4-8}&& $\bar{\nu}_\mu$ &-29.42&-0.08&-0.748&0.327&-0.087 \\ \midrule 
\multirow{2}{*}{5000} & \multirow{2}{*}{}& $\nu_\mu$ &-29.38&-0.192&-0.639&0.329&-0.105 \\ \cmidrule{4-8}&& $\bar{\nu}_\mu$ &-29.41&0.142&-1.11&0.529&-0.122 \\ \midrule 
\multirow{2}{*}{10000} & \multirow{2}{*}{}& $\nu_\mu$ &-29.32&-0.129&-0.787&0.415&-0.118 \\ \cmidrule{4-8}&& $\bar{\nu}_\mu$ &-29.32&-0.115&-0.682&0.286&-0.076 \\ \midrule 
\bottomrule 
\end{tabular} 
 \end{table} } 
 
{\renewcommand{\arraystretch}{0.3} 
 \begin{table}[htbp] 
 \footnotesize 
 \begin{tabular}{cccccccc} 
\toprule 
\multicolumn{8}{c}{Coupling to $s_R\bar{s}_R$} \\ [0.2cm] 
$m_{\mathrm{DM}}$ & $E_0$ & & $a_0 / b_0$ & $a_1 / b_1$ & $a_2 / b_2$ & $a_3 / b_3$ & $a_4 / b_4$\\[0.1cm] 
\toprule 
\multirow{2}{*}{100} & \multirow{2}{*}{35}& $\nu_\mu$ &-30.12/-33.62&-0.359/-22.59&-2.185/237.7&1.79/-1337&-0.809/2312 \\ \cmidrule{4-8}&& $\bar{\nu}_\mu$ &-30.09/-33.71&-0.586/-12.11&-1.326/55.64&0.877/-258.3&-0.517/342.2 \\ \midrule 
\multirow{2}{*}{200} & \multirow{2}{*}{100}& $\nu_\mu$ &-29.93/-34.99&-0.874/-8.105&-0.184/47.84&-0.181/-174.3&-0.082/34.2 \\ \cmidrule{4-8}&& $\bar{\nu}_\mu$ &-29.93/-34.96&-0.686/-3.935&-0.464/-5.125&-0.033/71.58&-0.103/-290.3 \\ \midrule 
\multirow{2}{*}{300} & \multirow{2}{*}{170}& $\nu_\mu$ &-29.83/-35.29&-1.073/-21.84&0.465/452.6&-0.676/-3843&0.074/9851 \\ \cmidrule{4-8}&& $\bar{\nu}_\mu$ &-29.83/-35.25&-0.967/-6.63&0.416/8.413&-0.743/364.1&0.112/-1950 \\ \midrule 
\multirow{2}{*}{500} & \multirow{2}{*}{}& $\nu_\mu$ &-29.71&-1.398&1.255&-1.171&0.194 \\ \cmidrule{4-8}&& $\bar{\nu}_\mu$ &-29.76&-0.766&0.136&-0.501&0.073 \\ \midrule 
\multirow{2}{*}{800} & \multirow{2}{*}{}& $\nu_\mu$ &-29.68&-1.039&0.554&-0.61&0.071 \\ \cmidrule{4-8}&& $\bar{\nu}_\mu$ &-29.71&-0.642&-0.044&-0.315&0.032 \\ \midrule 
\multirow{2}{*}{1000} & \multirow{2}{*}{}& $\nu_\mu$ &-29.64&-1.079&0.626&-0.615&0.069 \\ \cmidrule{4-8}&& $\bar{\nu}_\mu$ &-29.68&-0.661&0.022&-0.335&0.036 \\ \midrule 
\multirow{2}{*}{3000} & \multirow{2}{*}{}& $\nu_\mu$ &-29.56&-0.585&-0.166&-0.037&-0.047 \\ \cmidrule{4-8}&& $\bar{\nu}_\mu$ &-29.46&-1.178&0.875&-0.673&0.083 \\ \midrule 
\multirow{2}{*}{5000} & \multirow{2}{*}{}& $\nu_\mu$ &-29.49&-0.718&0.116&-0.19&-0.017 \\ \cmidrule{4-8}&& $\bar{\nu}_\mu$ &-29.52&-0.346&-0.429&0.064&-0.046 \\ \midrule 
\multirow{2}{*}{10000} & \multirow{2}{*}{}& $\nu_\mu$ &-29.44&-0.513&-0.196&0.02&-0.056 \\ \cmidrule{4-8}&& $\bar{\nu}_\mu$ &-29.42&-0.499&-0.155&-0.05&-0.029 \\ \midrule 
\bottomrule 
\end{tabular} 
 \end{table} } 
 
{\renewcommand{\arraystretch}{0.3} 
 \begin{table}[htbp] 
 \footnotesize 
 \begin{tabular}{cccccccc} 
\toprule 
\multicolumn{8}{c}{Coupling to $s_L\bar{s}_L$} \\ [0.2cm] 
$m_{\mathrm{DM}}$ & $E_0$ & & $a_0 / b_0$ & $a_1 / b_1$ & $a_2 / b_2$ & $a_3 / b_3$ & $a_4 / b_4$\\[0.1cm] 
\toprule 
\multirow{2}{*}{100} & \multirow{2}{*}{35}& $\nu_\mu$ &-30.08/-33.41&-0.927/3.269&-0.121/-60.42&-0.799/306.3&0.232/-570.2 \\ \cmidrule{4-8}&& $\bar{\nu}_\mu$ &-30.07/-33.35&-1.106/-2.25&0.826/40.88&-1.955/-225.6&0.64/267.3 \\ \midrule 
\multirow{2}{*}{200} & \multirow{2}{*}{100}& $\nu_\mu$ &-29.94/-33.27&-0.551/4.325&-0.846/-155.3&0.081/1025&0.039/-2250 \\ \cmidrule{4-8}&& $\bar{\nu}_\mu$ &-29.93/-33.18&-0.509/5.273&-0.666/-166.1&-0.208/1098&0.142/-2423 \\ \midrule 
\multirow{2}{*}{300} & \multirow{2}{*}{170}& $\nu_\mu$ &-29.87/-33.5&-0.37/-5.696&-1.191/-5.877&0.519/233.7&-0.103/-1260 \\ \cmidrule{4-8}&& $\bar{\nu}_\mu$ &-29.88/-33.44&-0.074/1.591&-1.624/-116.6&0.719/929.4&-0.125/-2661 \\ \midrule 
\multirow{2}{*}{500} & \multirow{2}{*}{}& $\nu_\mu$ &-29.88&0.499&-2.808&1.72&-0.383 \\ \cmidrule{4-8}&& $\bar{\nu}_\mu$ &-29.91&0.907&-3.409&2.011&-0.421 \\ \midrule 
\multirow{2}{*}{800} & \multirow{2}{*}{}& $\nu_\mu$ &-29.81&0.49&-2.546&1.51&-0.331 \\ \cmidrule{4-8}&& $\bar{\nu}_\mu$ &-29.84&0.867&-3.09&1.751&-0.355 \\ \midrule 
\multirow{2}{*}{1000} & \multirow{2}{*}{}& $\nu_\mu$ &-29.77&0.418&-2.261&1.283&-0.277 \\ \cmidrule{4-8}&& $\bar{\nu}_\mu$ &-29.8&0.847&-2.947&1.633&-0.325 \\ \midrule 
\multirow{2}{*}{3000} & \multirow{2}{*}{}& $\nu_\mu$ &-29.58&0.096&-1.472&0.797&-0.179 \\ \cmidrule{4-8}&& $\bar{\nu}_\mu$ &-29.57&0.129&-1.406&0.69&-0.143 \\ \midrule 
\multirow{2}{*}{5000} & \multirow{2}{*}{}& $\nu_\mu$ &-29.51&0.011&-1.244&0.661&-0.154 \\ \cmidrule{4-8}&& $\bar{\nu}_\mu$ &-29.53&0.317&-1.641&0.811&-0.162 \\ \midrule 
\multirow{2}{*}{10000} & \multirow{2}{*}{}& $\nu_\mu$ &-29.41&-0.153&-0.858&0.427&-0.111 \\ \cmidrule{4-8}&& $\bar{\nu}_\mu$ &-29.44&0.274&-1.501&0.726&-0.146 \\ \midrule 
\bottomrule 
\end{tabular} 
 \end{table} } 
 
 \FloatBarrier 
\section{Upper limits on the scattering cross sections for ${\rm DM}\,{\rm DM}\rightarrow Z Z$ and $gg$}
\label{ap:ZZ-gg}

As explained in this paper, the upper limit on the interaction cross section in scenarios where the dark matter annihilation is driven by a coupling to the electron, the muon or a light quark can be  approximately calculated from the corresponding upper limits assuming annihilations into $Z Z$ and $gg$. The 90\% C.L. limits on the spin-dependent and spin-independent interaction cross section for dark matter annihilating either to $ZZ$ or $gg$ are summarized in Table \ref{tab:VVchannels} as a function of the dark matter mass and shown in \Figref{fig:VVchannels}. The limits were obtained, following \cite{Ibarra:2013eba}, by choosing the cut angle between the reconstructed muon direction and the Sun that gives the best constraint under a background only hypothesis; the cut angle as a function of the dark matter mass is different for the $ZZ$ channel and for the $gg$ channel.

{\renewcommand{\arraystretch}{1.2}
\begin{table}[h!]
\begin{center}
\begin{tabular}{|c||c|c||c|c|}
\hline  
$m_{\text{DM}}[{\rm GeV}] $&
$\sigma_{\text{SD}}^{{\rm max}, ZZ} [{\rm cm}^2]$ & $\sigma_{\text{SI}}^{{\rm max}, ZZ} [{\rm cm}^2]$ & $\sigma_{\text{SD}}^{{\rm max}, gg} [{\rm cm}^2]$ & $\sigma_{\text{SI}}^{{\rm max}, gg} [{\rm cm}^2]$\\
\hline \hline
100	& $3.51 \cdot 10^{-40}$ & $7.88 \cdot 10^{-43}$ & $9.77 \cdot 10^{-37}$ & $2.19 \cdot 10^{-39}$\\ \hline
120	& $2.87 \cdot 10^{-40}$ & $5.65 \cdot 10^{-43}$ & $6.07 \cdot 10^{-37}$ & $1.19 \cdot 10^{-39}$\\ \hline
150	& $3.01 \cdot 10^{-40}$ & $5.10 \cdot 10^{-43}$ & $3.93 \cdot 10^{-37}$ & $6.65 \cdot 10^{-40}$\\ \hline
200	& $1.50 \cdot 10^{-40}$ & $2.12 \cdot 10^{-43}$ & $2.72 \cdot 10^{-37}$ & $3.85 \cdot 10^{-40}$\\ \hline
300	& $2.36 \cdot 10^{-40}$ & $2.69 \cdot 10^{-43}$ & $2.10 \cdot 10^{-37}$ & $2.39 \cdot 10^{-40}$\\ \hline
500	& $2.96 \cdot 10^{-40}$ & $2.73 \cdot 10^{-43}$ & $2.20 \cdot 10^{-37}$ & $2.03 \cdot 10^{-40}$\\ \hline
700	& $4.28 \cdot 10^{-40}$ & $3.57 \cdot 10^{-43}$ & $2.08 \cdot 10^{-37}$ & $1.73 \cdot 10^{-40}$\\ \hline
1000    & $7.10 \cdot 10^{-40}$ & $5.49 \cdot 10^{-43}$ & $3.56 \cdot 10^{-37}$ & $2.75 \cdot 10^{-40}$\\ \hline
2000	& $2.57 \cdot 10^{-39}$ & $1.81 \cdot 10^{-42}$ & $5.06 \cdot 10^{-37}$ & $3.57 \cdot 10^{-40}$\\ \hline
3000	& $6.17 \cdot 10^{-39}$ & $4.22 \cdot 10^{-42}$ & $6.84 \cdot 10^{-37}$ & $4.68 \cdot 10^{-40}$\\ \hline
5000	& $1.85 \cdot 10^{-38}$ & $1.24 \cdot 10^{-41}$ & $1.16 \cdot 10^{-36}$ & $7.76 \cdot 10^{-40}$\\ \hline
7000	& $3.80 \cdot 10^{-38}$ & $2.52 \cdot 10^{-41}$ & $1.65 \cdot 10^{-36}$ & $1.09 \cdot 10^{-39}$\\ \hline
10000	& $7.92 \cdot 10^{-38}$ & $5.21 \cdot 10^{-41}$ & $2.56 \cdot 10^{-36}$ & $1.68 \cdot 10^{-39}$ \\ \hline
\end{tabular}
\end{center}
\caption{\small 90\% C.L. limits on the spin-dependent and spin-independent interaction cross section for dark matter annihilating to $ZZ$ (first two columns) and $gg$ (last two columns), for various values of the dark matter mass $m_{\text{DM}}$.}
\label{tab:VVchannels}
\end{table}
}

\begin{figure}[h!]
\begin{center}
\includegraphics[width=0.49\textwidth]{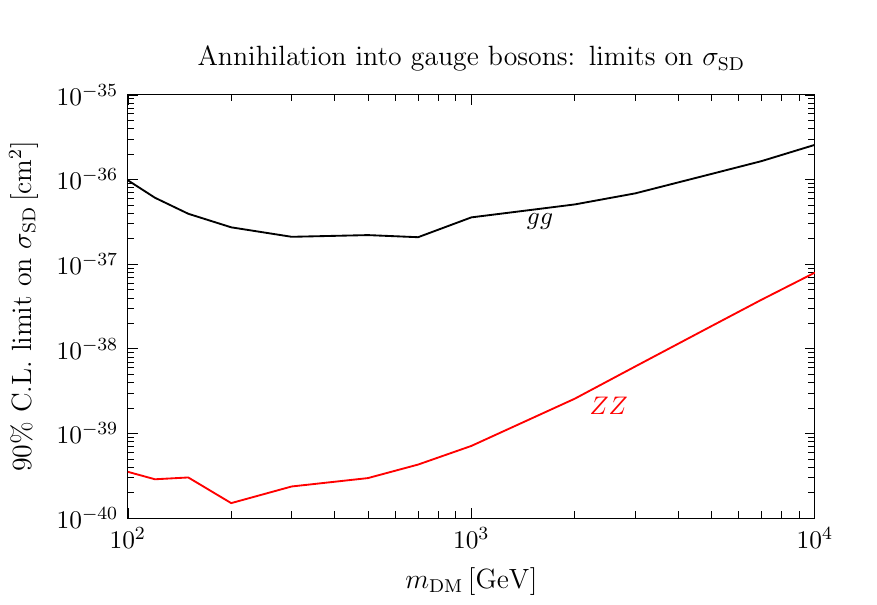}
\includegraphics[width=0.49\textwidth]{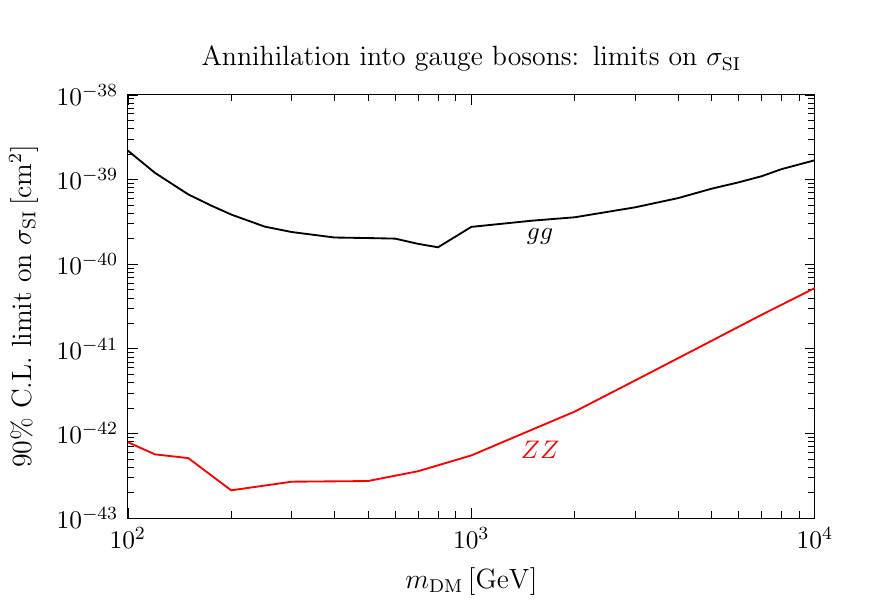}
\caption{\small 90\% C.L. limits on the spin-dependent (left plot) and spin-independent (right plot) interaction cross section for dark matter annihilating to $ZZ$ (red curve) and $gg$ (black curve).}
\label{fig:VVchannels}
\end{center}
\end{figure}

\section{Annihilation cross sections}
\label{ap:crosssections}
We include in this appendix the expressions for the relevant annihilation cross sections in the toy model discussed in Section \ref{sec:toy-model} in the limit $m_f\rightarrow 0$ and keeping the lowest order in the expansion in the relative dark matter velocity $v$. In the following formulas, $y$ is the Yukawa coupling between the Majorana dark matter particle $\chi$, the scalar $\eta$, and the right-handed Standard Model fermion $f_R$, which has electric charge $q_f$ and color $N_C$. The cross-sections for annihilations into $f_R\bar f_R$ were calculated in  \cite{Cao:2009yy}, for $\gamma \gamma$ and $g g$ via a one-loop diagram in \cite{Bergstrom:1997fh,Bern:1997ng}, for $\gamma Z$ in \cite{Ullio:1997ke}, for  $f_R\bar f_R \gamma$ in \cite{Bergstrom:1989jr,Flores:1989ru,Bringmann:2007nk}, for $f_R\bar f_R Z$ in\cite{Garny:2011ii} and for $f_R\bar f_R g$ in \cite{Flores:1989ru}; 
to the best of our knowledge,  the cross-section for annihilations into $ZZ$ at one loop is derived for the first time in this paper.\footnote{We have used FeynCalc~\cite{Mertig:1990an} for parts of the analytical computations.}

\subsubsection*{Two-to-two annihilation into fermions}
\label{app:TwoToTwoAnnihilation}
\begin{equation}
(\sigma v)_{f_R\bar f_R} = v^2\frac{y^4N_c}{48\pi m_\chi^2}\frac{1+m_{\eta}^4/m_\chi^4}{(1+m_{\eta}^2/m_\chi^2)^4} \;.
\end{equation}

\subsubsection*{Two-to-two annihilations into gauge bosons via loops}
\label{app:TwoToTwoLoopsAnnihilation}

\begin{align}
(\sigma v)_{\gamma \gamma} &= \frac{N_C^2 \, q_f^4 \, \alpha_{\text{em}}^2 \, y^4}{256 \pi^3 \, m_{\chi}^2} \left[ \text{Li}_2 \left( -\frac{m_{\chi}^2}{m_{\eta}^2} \right) - \text{Li}_2 \left( \frac{m_{\chi}^2}{m_{\eta}^2} \right)\right]^2 \;,\\
(\sigma v)_{gg} &= \frac{2 \, \alpha_{\text{s}}^2 \, y^4}{256 \pi^3 \, m_{\chi}^2} \left[ \text{Li}_2 \left( -\frac{m_{\chi}^2}{m_{\eta}^2} \right) - \text{Li}_2 \left( \frac{m_{\chi}^2}{m_{\eta}^2} \right)\right]^2 \;,\\
(\sigma v)_{\gamma \text{Z}} &= \frac{\left| \mathcal{A}_{\gamma \text{Z}} \right|^2}{512 \, \pi^3 \, m_{\chi}^6 \, m_{\eta}^4 \left(1-\frac{m_{\text{Z}}^2}{4 \, m_{\chi}^2} \right) \left( 1-\frac{m_{\text{Z}}^4}{16 \, m_{\eta}^4} \right)^2}\;, \\
(\sigma v)_{\text{Z} \text{Z}} &= \frac{\left| \mathcal{A}_{\text{Z} \text{Z}} \right|^2}{1024 \, \pi^3 \, m_{\chi}^6 \, m_{\eta}^4 \, \sqrt{1-\frac{m_{\text{Z}}^2}{m_{\chi}^2}}}\;,
\end{align}
with ${\rm Li}_2(x)$ the dilogarithm function, while $\mathcal{A}_{\gamma \text{Z}}$ and  $\mathcal{A}_{\text{Z} \text{Z}}$ are defined by
\begin{align}
\mathcal{A}_{\gamma \text{Z}} &= N_C \, q_f^2 \, \alpha_{\text{em}} \, y^2 \, \tan \left( \theta_{\text{W}} \right) \left(1-\frac{m_{\text{Z}}^2}{4 \,m_{\eta}^2}\right) \Bigg\{ \nonumber \\
& m_{\text{Z}}^2 \left(\frac{m_{\chi}^4+m_{\eta}^4}{2}+\frac{m_{\text{Z}}^2 \left(m_{\eta}^2-m_{\chi}^2\right)}{4}+\frac{m_{\text{Z}}^4}{16}\right) \, \, \text{C}_0 \left(m_{\chi}^2,m_{\text{Z}}^2,\frac{m_{\text{Z}}^2}{2}-m_{\chi}^2,m_{\eta}^2,0,0\right) \nonumber \\
& + 2 \, m_{\eta}^2 \left(m_{\chi}^2-\frac{m_{\text{Z}}^2}{4}\right)^2 \text{C}_0 \left(m_{\chi}^2,0,\frac{m_{\text{Z}}^2}{2}-m_{\chi}^2,0,m_{\eta}^2,m_{\eta}^2\right) \nonumber\\
& +\left(m_{\eta}^2+\frac{m_{\text{Z}}^2}{4}\right) \left(2 m_{\chi}^4-m_{\chi}^2 m_{\text{Z}}^2+\frac{m_{\eta}^2 m_{\text{Z}}^2}{2}\right) \text{C}_0 \left(m_{\chi}^2,m_{\text{Z}}^2,\frac{m_{\text{Z}}^2}{2}-m_{\chi}^2,0,m_{\eta}^2,m_{\eta}^2\right) \nonumber\\
& + \, \frac{m_{\text{Z}}^2}{2} \left(m_{\eta}^2+\frac{m_{\text{Z}}^2}{4}\right) \, \, \left[2 \, \sqrt{\frac{4 m_{\eta}^2}{m_{\text{Z}}^2}-1} \, \, \text{arccot} \left(\sqrt{\frac{4 m_{\eta}^2}{m_{\text{Z}}^2}-1}\right)-\log \frac{m_{\text{Z}}^2}{m_{\eta}^2}+i \pi \right] \Bigg\} \, ,
\end{align}
\begin{align}
\mathcal{A}_{\text{ZZ}} &=  N_C \, q_f^2 \, \alpha_{\text{em}} \, y^2 \, \tan^2 \left( \theta_{\text{W}} \right) \Bigg\{ m_{\text{Z}}^2 \left(m_{\chi}^4+m_{\eta}^4-m_{\chi}^2 m_{\text{Z}}^2\right) \text{C}_0 \left(m_{\chi}^2,m_{\text{Z}}^2,-m_{\chi}^2+m_{\text{Z}}^2,m_{\eta}^2,0,0\right) \nonumber \\
& +\left[4 m_{\chi}^4 m_{\eta}^2+\left(-m_{\chi}^4-4 m_{\chi}^2 m_{\eta}^2+m_{\eta}^4\right) m_{\text{Z}}^2+m_{\chi}^2 m_{\text{Z}}^4\right] \text{C}_0 \left(m_{\chi}^2,m_{\text{Z}}^2,-m_{\chi}^2+m_{\text{Z}}^2,0,m_{\eta}^2,m_{\eta}^2\right) \nonumber \\
& + m_{\eta}^2 \, m_{\text{Z}}^2 \left[2 \sqrt{\frac{4 m_{\eta}^2}{m_{\text{Z}}^2}-1} \, \, \text{arccot} \left(\sqrt{\frac{4 m_{\eta}^2}{m_{\text{Z}}^2}-1}\right)-\text{log} \, \frac{m_{\text{Z}}^2}{m_{\eta}^2} +i \pi \right] \Bigg\} \, ,
\end{align}
$C_0$ being a Passarino-Veltman function. These expressions satisfy 
\begin{align}
\left( \sigma v \right)_{\text{Z} \text{Z}} \big|_{\tan \left( \theta_{\text{W}} \right) \equiv 1} \stackrel{m_{\text{Z}} \rightarrow 0}{\longrightarrow} \left( \sigma v \right)_{\gamma \gamma},~~~~~
\left( \sigma v \right)_{\gamma \text{Z}} \big|_{\tan \left( \theta_{\text{W}} \right) \equiv 1} \stackrel{m_{\text{Z}} \rightarrow 0}{\longrightarrow} 2 \, \left( \sigma v \right)_{\gamma \gamma} \,.
\end{align}

\subsubsection*{Two-to-three annihilations}
\label{app:TwoToThreeAnnihilations}

\begin{eqnarray}
\frac{d(\sigma v)_{f_R\bar f_R \gamma}}{dE_\gamma dE_f} & = & \frac{q_f^2 N_C\alpha_{em}y^4 \left(1-\frac{E_\gamma}{m_\chi}\right) \left[\left(\frac{E_\gamma}{m_\chi}\right)^2-2\frac{E_\gamma}{m_\chi} \left((1-\frac{E_f}{m_\chi}\right)+2 \left(1-\frac{E_f}{m_\chi}\right)^2 \right]}{8\pi^2 m_\chi^4 \left(1-2\frac{E_f}{m_\chi}-\frac{m_\eta^2}{m_\chi^2}\right)^2 \left(3-2\frac{E_\gamma}{m_\chi}-2\frac{E_f}{m_\chi}+\frac{m_\eta^2}{m_\chi^2}\right)^2} \;, \\
\frac{d(\sigma v)_{f_R\bar f_R Z}}{dE_Z dE_f} & = & \frac{q_f^2N_c\tan^2(\theta_W)\alpha_{em}y^4 }{8\pi^2 m_\chi^4 \left(1-2\frac{E_f}{m_\chi}-\frac{m_\eta^2}{m_\chi^2}\right)^2 \left(3-2x-2\frac{E_f}{m_\chi}+\frac{m_\eta^2}{m_\chi^2}\right)^2}  \nonumber \\
&& {} \times \Big\{ \left(1-\frac{E_Z}{m_\chi}\right)\left[\left(\frac{E_Z}{m_\chi}\right)^2-2\frac{E_Z}{m_\chi} \left(1-\frac{E_f}{m_\chi}\right)+2 \left(1-\frac{E_f}{m_\chi}\right)^2 \right] \nonumber \\
&& {} + \left(\frac{m_Z}{m_\chi}\right)^2 \left[\left(\frac{E_Z}{m_\chi}\right)^2+2\left(\frac{E_f}{m_\chi}\right)^2+2\frac{E_Z}{m_\chi}\frac{E_f}{m_\chi} -4\frac{E_f}{m_\chi}\right]/4 \nonumber \\
&&{} -\left(\frac{m_Z}{m_\chi}\right)^4/8 \Big \} \;, \\
\frac{d(\sigma v)_{f_R\bar f_R g}}{dE_\gamma dE_f} & = & \frac{\left(N_c^2-1\right)\alpha_{s}(m_\chi)y^4}{16\pi^2 m_\chi^4 (1-2\frac{E_f}{m_\chi}-\frac{m_\eta^2}{m_\chi^2})^2\left(3-2\frac{E_g}{m_\chi}-2\frac{E_f}{m_\chi}+\frac{m_\eta^2}{m_\chi^2}\right)^2} \nonumber \\
&&{} \times  \left(1-\frac{E_g}{m_\chi}\right) \left[\left(\frac{E_g}{m_\chi}\right)^2-2\frac{E_g}{m_\chi}(1-\frac{E_f}{m_\chi})+2\left(1-\frac{E_f}{m_\chi}\right)^2 \right]\;.
\end{eqnarray}
The spectra of gauge bosons are obtained by integrating the differential cross-section over the fermion energy, with integration limits given by $E_f^{\rm min/max}=m_\chi-(E_V\pm\sqrt{E_V^2-M_V^2})/2$. The total cross-section can be obtained by integrating over the remaining energy with limits $E_V^{\rm min}=M_V$ and $E_V^{\rm max}=m_\chi+M_V^2/(4m_\chi)$. 

\newpage

%%%%%%%%%%%%%%%%%


\bibliographystyle{JHEP-mod}
%\bibliography{../../archiv}
\bibliography{ITW}

\providecommand{\href}[2]{#2}\begingroup\raggedright\begin{thebibliography}{10}

\bibitem{Aartsen:2012kia}
{\bf IceCube collaboration}, M.~Aartsen {\em et.~al.}, {\it {Search for dark
  matter annihilations in the Sun with the 79-string IceCube detector}},  {\em
  Phys.Rev.Lett.} {\bf 110} (2013) 131302,
  [\href{http://xxx.lanl.gov/abs/1212.4097}{{\tt arXiv:1212.4097}}].

\bibitem{Behnke:2012ys}
{\bf COUPP Collaboration}, E.~Behnke {\em et.~al.}, {\it {First Dark Matter
  Search Results from a 4-kg CF$_3$I Bubble Chamber Operated in a Deep
  Underground Site}},  {\em Phys.Rev.} {\bf D86} (2012) 052001,
  [\href{http://xxx.lanl.gov/abs/1204.3094}{{\tt arXiv:1204.3094}}].

\bibitem{Felizardo:2011uw}
M.~Felizardo, T.~Girard, T.~Morlat, A.~Fernandes, A.~Ramos, {\em et.~al.}, {\it
  {Final Analysis and Results of the Phase II SIMPLE Dark Matter Search}},
  {\em Phys.Rev.Lett.} {\bf 108} (2012) 201302,
  [\href{http://xxx.lanl.gov/abs/1106.3014}{{\tt arXiv:1106.3014}}].

\bibitem{Bernal:2012qh}
N.~Bernal, J.~Martin-Albo, and S.~Palomares-Ruiz, {\it {A novel way of
  constraining WIMPs annihilations in the Sun: MeV neutrinos}},  {\em JCAP}
  {\bf 1308} (2013) 011, [\href{http://xxx.lanl.gov/abs/1208.0834}{{\tt
  arXiv:1208.0834}}].

\bibitem{Rott:2012qb}
C.~Rott, J.~Siegal-Gaskins, and J.~F. Beacom, {\it {New Sensitivity to Solar
  WIMP Annihilation using Low-Energy Neutrinos}},  {\em Phys.Rev.} {\bf D88}
  (2013) 055005, [\href{http://xxx.lanl.gov/abs/1208.0827}{{\tt
  arXiv:1208.0827}}].

\bibitem{Kachelriess:2009zy}
M.~Kachelriess, P.~Serpico, and M.~A. Solberg, {\it {On the role of electroweak
  bremsstrahlung for indirect dark matter signatures}},  {\em Phys.Rev.} {\bf
  D80} (2009) 123533, [\href{http://xxx.lanl.gov/abs/0911.0001}{{\tt
  arXiv:0911.0001}}].

\bibitem{Ciafaloni:2010ti}
P.~Ciafaloni, D.~Comelli, A.~Riotto, F.~Sala, A.~Strumia, {\em et.~al.}, {\it
  {Weak Corrections are Relevant for Dark Matter Indirect Detection}},  {\em
  JCAP} {\bf 1103} (2011) 019, [\href{http://xxx.lanl.gov/abs/1009.0224}{{\tt
  arXiv:1009.0224}}].

\bibitem{Griest:1986yu}
K.~Griest and D.~Seckel, {\it {Cosmic Asymmetry, Neutrinos and the Sun}},  {\em
  Nucl.Phys.} {\bf B283} (1987) 681.

\bibitem{Gould:1987ir}
A.~Gould, {\it {Resonant Enhancements in WIMP Capture by the Earth}},  {\em
  Astrophys.J.} {\bf 321} (1987) 571.

\bibitem{Gondolo:2004sc}
P.~Gondolo, J.~Edsjo, P.~Ullio, L.~Bergstrom, M.~Schelke, {\em et.~al.}, {\it
  {DarkSUSY: Computing supersymmetric dark matter properties numerically}},
  {\em JCAP} {\bf 0407} (2004) 008,
  [\href{http://xxx.lanl.gov/abs/astro-ph/0406204}{{\tt astro-ph/0406204}}].

\bibitem{Green:2011bv}
A.~M. Green, {\it {Astrophysical uncertainties on direct detection
  experiments}},  {\em Mod.Phys.Lett.} {\bf A27} (2012) 1230004,
  [\href{http://xxx.lanl.gov/abs/1112.0524}{{\tt arXiv:1112.0524}}].

\bibitem{Ciafaloni:2001mu}
M.~Ciafaloni, P.~Ciafaloni, and D.~Comelli, {\it {Towards collinear evolution
  equations in electroweak theory}},  {\em Phys.Rev.Lett.} {\bf 88} (2002)
  102001, [\href{http://xxx.lanl.gov/abs/hep-ph/0111109}{{\tt
  hep-ph/0111109}}].

\bibitem{Ciafaloni:2005fm}
P.~Ciafaloni and D.~Comelli, {\it {Electroweak evolution equations}},  {\em
  JHEP} {\bf 0511} (2005) 022,
  [\href{http://xxx.lanl.gov/abs/hep-ph/0505047}{{\tt hep-ph/0505047}}].

\bibitem{Altarelli:1977zs}
G.~Altarelli and G.~Parisi, {\it {Asymptotic Freedom in Parton Language}},
  {\em Nucl.Phys.} {\bf B126} (1977) 298.

\bibitem{Sjostrand:2006za}
T.~Sjostrand, S.~Mrenna, and P.~Z. Skands, {\it {PYTHIA 6.4 Physics and
  Manual}},  {\em JHEP} {\bf 0605} (2006) 026,
  [\href{http://xxx.lanl.gov/abs/hep-ph/0603175}{{\tt hep-ph/0603175}}].

\bibitem{Sjostrand:2007gs}
T.~Sjostrand, S.~Mrenna, and P.~Z. Skands, {\it {A Brief Introduction to PYTHIA
  8.1}},  {\em Comput.Phys.Commun.} {\bf 178} (2008) 852--867,
  [\href{http://xxx.lanl.gov/abs/0710.3820}{{\tt arXiv:0710.3820}}].

\bibitem{Christiansen:2014kba}
J.~R. Christiansen and T.~Sjostrand, {\it {Weak Gauge Boson Radiation in Parton
  Showers}},  \href{http://xxx.lanl.gov/abs/1401.5238}{{\tt arXiv:1401.5238}}.

\bibitem{Baratella:2013fya}
P.~Baratella, M.~Cirelli, A.~Hektor, J.~Pata, M.~Piibeleht, {\em et.~al.}, {\it
  {PPPC 4 DM$\nu$: A Poor Particle Physicist Cookbook for Neutrinos from DM
  annihilations in the Sun}},  \href{http://xxx.lanl.gov/abs/1312.6408}{{\tt
  arXiv:1312.6408}}.

\bibitem{Ibarra:2013eba}
A.~Ibarra, M.~Totzauer, and S.~Wild, {\it {High-energy neutrino signals from
  the Sun in dark matter scenarios with internal bremsstrahlung}},  {\em JCAP}
  {\bf 1312} (2013) 043, [\href{http://xxx.lanl.gov/abs/1311.1418}{{\tt
  arXiv:1311.1418}}].

\bibitem{Blennow:2007tw}
M.~Blennow, J.~Edsjo, and T.~Ohlsson, {\it {Neutrinos from WIMP annihilations
  using a full three-flavor Monte Carlo}},  {\em JCAP} {\bf 0801} (2008) 021,
  [\href{http://xxx.lanl.gov/abs/0709.3898}{{\tt arXiv:0709.3898}}].

\bibitem{GonzalezGarcia:2012sz}
M.~Gonzalez-Garcia, M.~Maltoni, J.~Salvado, and T.~Schwetz, {\it {Global fit to
  three neutrino mixing: critical look at present precision}},  {\em JHEP} {\bf
  1212} (2012) 123, [\href{http://xxx.lanl.gov/abs/1209.3023}{{\tt
  arXiv:1209.3023}}].

\bibitem{Scott:2012mq}
{\bf IceCube Collaboration}, P.~Scott {\em et.~al.}, {\it {Use of event-level
  neutrino telescope data in global fits for theories of new physics}},  {\em
  JCAP} {\bf 1211} (2012) 057, [\href{http://xxx.lanl.gov/abs/1207.0810}{{\tt
  arXiv:1207.0810}}].

\bibitem{DanningerPhD}
M. Danninger, PhD thesis, Stockholm University.

\bibitem{Aprile:2012nq}
{\bf XENON100 Collaboration}, E.~Aprile {\em et.~al.}, {\it {Dark Matter
  Results from 225 Live Days of XENON100 Data}},  {\em Phys.Rev.Lett.} {\bf
  109} (2012) 181301, [\href{http://xxx.lanl.gov/abs/1207.5988}{{\tt
  arXiv:1207.5988}}].

\bibitem{Akerib:2013tjd}
{\bf LUX Collaboration}, D.~Akerib {\em et.~al.}, {\it {First results from the
  LUX dark matter experiment at the Sanford Underground Research Facility}},
  \href{http://xxx.lanl.gov/abs/1310.8214}{{\tt arXiv:1310.8214}}.

\bibitem{Chen:2013gya}
J.-Y. Chen, E.~W. Kolb, and L.-T. Wang, {\it {Dark matter coupling to
  electroweak gauge and Higgs bosons: an effective field theory approach}},
  \href{http://xxx.lanl.gov/abs/1305.0021}{{\tt arXiv:1305.0021}}.

\bibitem{Pukhov:1999gg}
A.~Pukhov, E.~Boos, M.~Dubinin, V.~Edneral, V.~Ilyin, {\em et.~al.}, {\it
  {CompHEP: A Package for evaluation of Feynman diagrams and integration over
  multiparticle phase space}},
  \href{http://xxx.lanl.gov/abs/hep-ph/9908288}{{\tt hep-ph/9908288}}.

\bibitem{Pukhov:2004ca}
A.~Pukhov, {\it {CalcHEP 2.3: MSSM, structure functions, event generation,
  batchs, and generation of matrix elements for other packages}},
  \href{http://xxx.lanl.gov/abs/hep-ph/0412191}{{\tt hep-ph/0412191}}.

\bibitem{Ritz:1987mh}
S.~Ritz and D.~Seckel, {\it {Detailed Neutrino Spectra From Cold Dark Matter
  Annihilations in the Sun}},  {\em Nucl.Phys.} {\bf B304} (1988) 877.

\bibitem{Cao:2009yy}
Q.-H. Cao, E.~Ma, and G.~Shaughnessy, {\it {Dark Matter: The Leptonic
  Connection}},  {\em Phys.Lett.} {\bf B673} (2009) 152--155,
  [\href{http://xxx.lanl.gov/abs/0901.1334}{{\tt arXiv:0901.1334}}].

\bibitem{Bergstrom:1997fh}
L.~Bergstrom and P.~Ullio, {\it {Full one loop calculation of neutralino
  annihilation into two photons}},  {\em Nucl.Phys.} {\bf B504} (1997) 27--44,
  [\href{http://xxx.lanl.gov/abs/hep-ph/9706232}{{\tt hep-ph/9706232}}].

\bibitem{Bern:1997ng}
Z.~Bern, P.~Gondolo, and M.~Perelstein, {\it {Neutralino annihilation into two
  photons}},  {\em Phys.Lett.} {\bf B411} (1997) 86--96,
  [\href{http://xxx.lanl.gov/abs/hep-ph/9706538}{{\tt hep-ph/9706538}}].

\bibitem{Ullio:1997ke}
P.~Ullio and L.~Bergstrom, {\it {Neutralino annihilation into a photon and a Z
  boson}},  {\em Phys.Rev.} {\bf D57} (1998) 1962--1971,
  [\href{http://xxx.lanl.gov/abs/hep-ph/9707333}{{\tt hep-ph/9707333}}].

\bibitem{Bergstrom:1989jr}
L.~Bergstrom, {\it {Radiative Processes in Dark Matter Photino Annihilation}},
  {\em Phys.Lett.} {\bf B225} (1989) 372.

\bibitem{Flores:1989ru}
R.~Flores, K.~A. Olive, and S.~Rudaz, {\it {Radiative Processes in Lsp
  Annihilation}},  {\em Phys.Lett.} {\bf B232} (1989) 377--382.

\bibitem{Bringmann:2007nk}
T.~Bringmann, L.~Bergstrom, and J.~Edsjo, {\it {New Gamma-Ray Contributions to
  Supersymmetric Dark Matter Annihilation}},  {\em JHEP} {\bf 0801} (2008) 049,
  [\href{http://xxx.lanl.gov/abs/0710.3169}{{\tt arXiv:0710.3169}}].

\bibitem{Garny:2011ii}
M.~Garny, A.~Ibarra, and S.~Vogl, {\it {Dark matter annihilations into two
  light fermions and one gauge boson: General analysis and antiproton
  constraints}},  {\em JCAP} {\bf 1204} (2012) 033,
  [\href{http://xxx.lanl.gov/abs/1112.5155}{{\tt arXiv:1112.5155}}].

\bibitem{Mertig:1990an}
R.~Mertig, M.~Bohm, and A.~Denner, {\it {FEYN CALC: Computer algebraic
  calculation of Feynman amplitudes}},  {\em Comput.Phys.Commun.} {\bf 64}
  (1991) 345--359.

\end{thebibliography}\endgroup

\end{document}